\RequirePackage{lineno}%DELETE FOR PLAIN VERSION
\documentclass[superscriptaddress,aps,preprint,amsmath,amssymb,floatfix]{revtex4-1}

\usepackage{graphicx,morefloats}
\usepackage{subfigure} % referring to Figure a and b separately
\usepackage{color,soul}
\usepackage{bm}
\usepackage{longtable}

\usepackage{amsmath}
\usepackage{amssymb}
\usepackage{times}
\usepackage{hyperref}
\usepackage{dsfont}
\usepackage{bbm}
\usepackage[normalem]{ulem}
\usepackage{gensymb}

\newcommand{\be}[0]{\begin{equation}}
\newcommand{\ee}[0]{\end{equation}}
\newcommand{\ba}[0]{\begin{eqnarray}}
\newcommand{\ea}[0]{\end{eqnarray}}
\newcommand{\mx}[0]{\begin{pmatrix}}
\newcommand{\ex}[0]{\end{pmatrix}}

\newcommand{\kbf}{\ensuremath{{\bf k}}}

\newcommand{\up}[0]{\uparrow}
\newcommand{\dn}[0]{\downarrow}

\newcommand{\lcr}[1]{{\color{black} #1}}

\begin{document}
% \linenumbers
%%%%%% USE: Ref.\citenum{xxx} INSTEAD OF Ref.\,\onlinecite{xxx} %%%%%

\hyphenation{va-ni-sh-ing}

\begin{center}

\thispagestyle{empty}

{\large\bf Emergent flat band and topological Kondo semimetal}\\
{\large\bf 
driven by orbital-selective correlations}
\\%[0.6cm]
[0.3cm]

Lei\ Chen$^{1}$,
Fang\ Xie$^{1}$,  Shouvik\ Sur$^1$, 
Haoyu\ Hu$^{1,2}$, \linebreak
Silke\  Paschen$^{3,1}$, 
Jennifer Cano$^{4,5}$, 
and Qimiao Si$^{1,\ast}$
\\[0.3cm]

$^1$Department of Physics and Astronomy, Rice Center for Quantum Materials, Rice University, Houston, Texas 77005, USA\\[-0.cm]

$^2$Donostia International Physics Center, P. Manuel de Lardizabal 4, 20018 Donostia-San Sebastian, Spain\\[-0.cm]

$^3$Institute of Solid State Physics, Vienna University of Technology, Wiedner Hauptstr. 8-10, 1040
Vienna, Austria\\[-0.0cm]

$^4$Department of Physics and Astronomy, Stony Brook University, Stony Brook, NY 11794, USA\\[-0.cm]

$^5$Center for Computational Quantum Physics, Flatiron Institute, New York, NY 10010, USA

\normalsize{$^\ast$To whom correspondence should be addressed; E-mail:  qmsi@rice.edu.}

\end{center}

%\vspace{0.3cm}
\vspace{0.16cm}

\noindent
{\large \bf Abstract }

\noindent
{\bf 
Flat electronic bands are expected to show proportionally enhanced electron correlations, 
which may generate a plethora of novel quantum phases and unusual low-energy excitations. 
They are increasingly being pursued in 
$d$-electron-based systems with crystalline lattices that feature 
destructive electronic interference, 
where they are often topological. 
Such flat bands, though, are generically located far away from the Fermi energy, which limits their capacity to partake in the low-energy physics. Here we show that electron correlations produce emergent flat bands that are pinned to the Fermi energy. We demonstrate this effect within a Hubbard model, in the regime described by Wannier orbitals
where an effective Kondo description arises through
orbital-selective Mott correlations.
Moreover, the correlation effect cooperates with symmetry constraints to produce 
a topological Kondo semimetal.
Our results motivate a novel design principle for
Weyl Kondo semimetals in a
new setting, viz. $d$-electron-based 
materials
on suitable crystal lattices,
and uncover 
interconnections among seemingly disparate systems
that may inspire
fresh understandings and realizations 
of correlated topological effects in quantum materials and beyond.
}

\clearpage
\newpage

\noindent
{\large \bf Introduction }

\noindent
Certain crystalline lattices feature flat bands, 
via frustration caused by destructive interference in electron motion \cite{Mielke1991},
 which are increasingly being explored
 in $d$-electron-based systems \cite{Ye2018,Yao18.1x}.
The reduced bandwidth correspondingly
enhances the effect of electron correlations. In addition, 
such 
flat bands are often topologically nontrivial.
As such, these systems represent a new platform to uncover novel physics for both correlation and topology as well as their interplay \cite{Pas21.1}.
For example, kagome metals may host flat bands and have  been the subject of considerable recent interest 
for realizing unusual forms of
charge-density-wave order
\cite{Hasan2020,Guguchia2022-TRSB,Zhou2021.2,Dai2022,Yin2022,Setty2022}.
They have also been implicated to exhibit 
a type of strange metal behavior \cite{Hu-flat-wide22.1,Huang2023np,Ye2021.x,Ekahana2021.x}
that resembles what has been 
extensively studied
in quantum critical heavy fermion metals
\cite{Pas21.1,Wir16.1,Kirchner2020}.

In order to strongly influence 
the low-energy physics, the flat bands need to be placed near the Fermi energy. 
However, 
this typically is not the case
at the level of bare (noninteracting) electron bandstructure.
There have been considerable recent experimental 
efforts to tune 
the bare flat bands 
to the vicinity of the Fermi energy.
With the rare exception coming from materials search \cite{Ye2021.x},
the tuning study
has met 
with only a limited success \cite{Zhang-fb-kagome2021,Deng-FGT2018}.
Because the flat bands are associated with $d$-electrons, the energy scales that determine the 
flat-band placement (with respect to the Fermi energy)
are relatively
large and, as a result, it is challenging to achieve
the required tuning.
We are thus motivated to ask the following 
important questions: Can electron correlations generate
emergent flat bands 
at
the Fermi energy in $d$-electron-based systems? 
And, if so, 
to what extent do the resulting phases display nontrivial correlation and/or topological physics?

\begin{figure}[h]
\centering
\includegraphics[width=0.9\columnwidth]{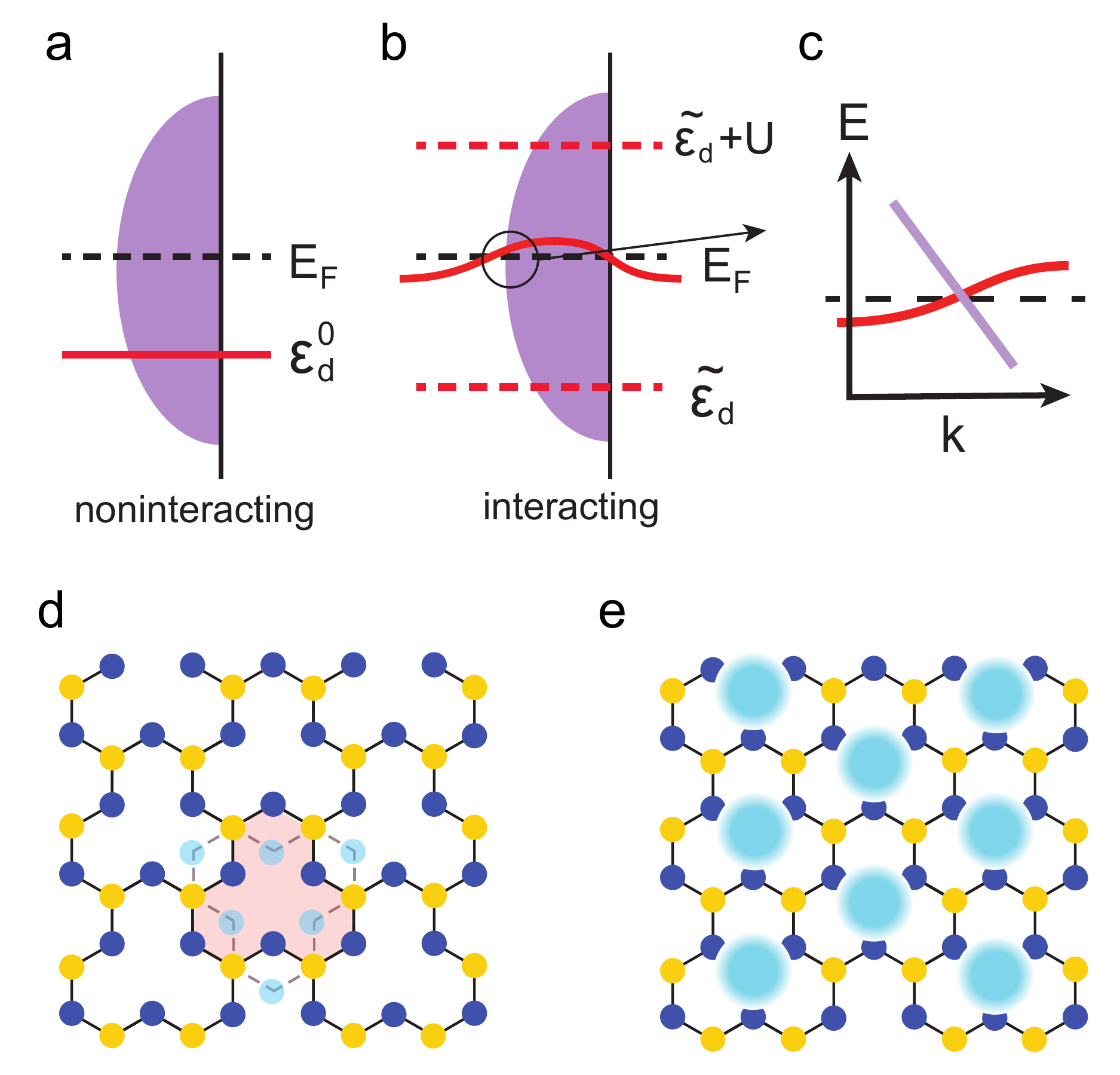}
\caption{\textbf{Illustration of the bare and emergent flat bands and lattice geometry.} {\bf a,}
In the noninteracting case, a flat band (red solid line) appears far away from the Fermi energy. 
{\bf b,}
In the presence of orbital-selective correlations, an interaction-driven flat band emerges at the Fermi energy (red solid line), while leaving incoherent excitations far away from the Fermi energy (red dashed lines). 
{\bf c,}
The emergent flat band crosses a dispersive band, leading to a topological Kondo semimetal with symmetry-protected Dirac/Weyl nodes that are pinned close to the Fermi energy, within an effective Kondo energy scale. {\bf d,}
 Geometry of the clover lattice with 5 sublattices per unit cell. The lattice does not have inversion symmetry. This can be seen from the mismatch between the (dark) blue sublattices and their inversion counterparts (dots in light blue).
 {\bf e,}
The 
Wannier orbitals are near the geometric centers (shaded blue circles) of the unit cells,  which form
 a triangular lattice.
}
\label{fig:sch}
\end{figure}

We address both issues in a Hubbard model in which the noninteracting limit features a topologically nontrivial flat band that is far away from the Fermi energy. 
Due to a 
well-separated hierarchy in
the widths
of the flat band and wide bands that it 
is coupled to,
and through the formation of compact molecular orbitals \cite{Chen2023-kagome},
orbital-selective Mott correlations develop 
\cite{Hu-flat-wide22.1}.
We show that such orbital-selective
correlations lead to emergent flat bands that are pinned to the Fermi energy. Moreover, 
using symmetry constraints in interacting settings \cite{Hu2021},
which are based on Green's function eigenstates (as opposed to 
Bloch states \cite{Bradlyn2017,Cano2018,Po2017,Watanabe2017,Cano2021}), 
we demonstrate that
the emergent flat bands 
lead to a topological Kondo semimetal. The latter is in the same family as Weyl Kondo semimetals that appear
 in topological Kondo-lattice models 
\cite{Lai2018, Grefe-prb20.1} and  materials 
\cite{Dzsaber2017,Dzs-giant21.1}
of both 
existing and 
designed \cite{Chen-Natphys22,Hu2021}
heavy fermion
systems.
The qualitative physics is illustrated in 
Figs.\,\ref{fig:sch}a-c. Importantly, our approach is based on an exact construction of molecular orbitals and an exact mapping to a heavy fermion description; this is in addition to the exact constraints that symmetry places, which will also come into our analysis.
Our results motivate a design principle for 
the Weyl Kondo semimetals in the new setting of 
$d$-electron-based systems,
and 
point to  the realization of
fractional Chern insulators \cite{Xie-fci2021}
in 
transition-metal compounds.

\noindent
{\large \bf Results }

\noindent
{\bf One-orbital Hubbard model on the clover lattice~~}

\begin{figure}[h]
\centering
\includegraphics[width=0.9\columnwidth]{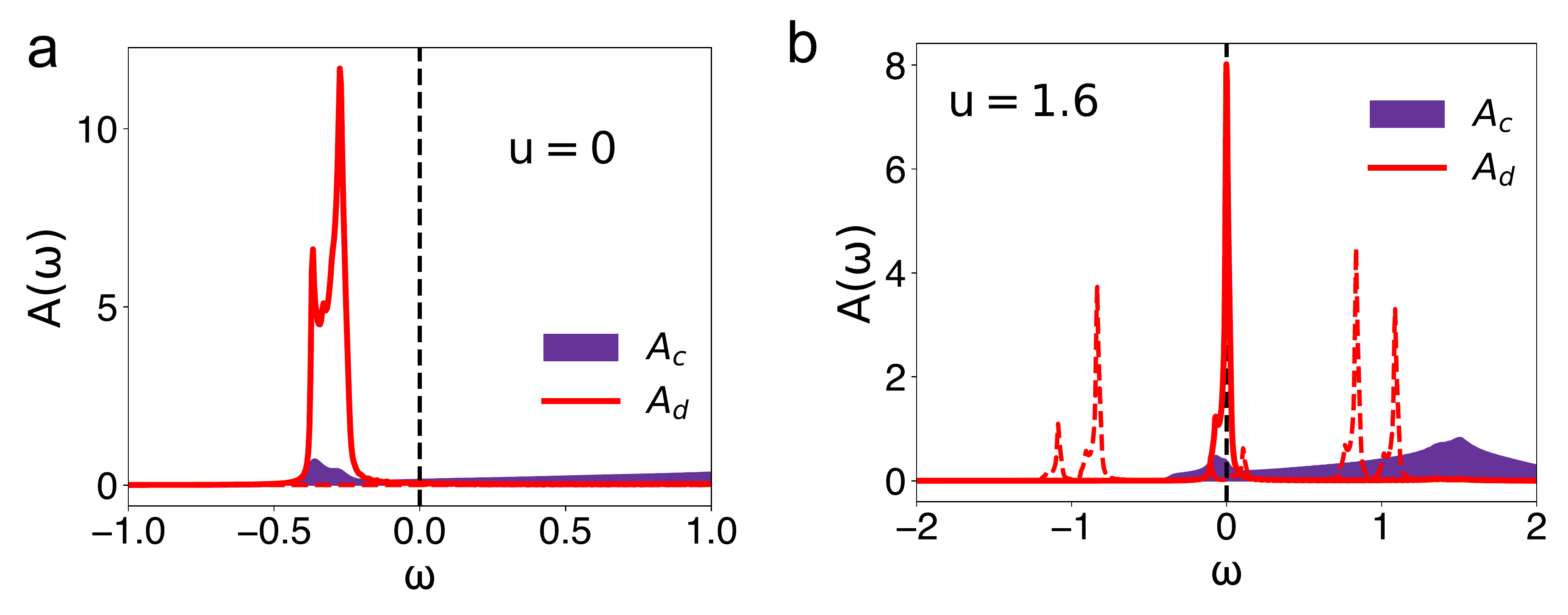}
\caption{
\textbf{Bare and emergent flat bands.}
{\bf a,}
The DOS of the $d$ and $c$ electrons for the noninteracting case.
{\bf b,} The corresponding result at $u=1.6$; here
the red solid (dashed) lines denote the coherent (incoherent) part of the $d$-electron excitations. The purple backgrounds mark the DOS of the conduction $c$ electrons.
}
\label{fig:dos}
\end{figure}

\noindent
For a proof-of-principle demonstration, we consider
a variant of the kagome lattice, the two-dimensional (2D) clover lattice.
As shown in Fig.\,\ref{fig:sch}d, 
it
contains five sublattices per unit cell. 
Leaving the details of the model 
to be given in the Methods and in 
Supplementary Note\,1, we note that this
lattice features a flat band (Supplementary Note\,2).
As a case study, 
the model is simplified while preserving the topological nature of the flat band; we do so by removing the $C_{3}$ symmetry of the clover lattice
 (Methods), leaving only a mirror symmetry $M_x$.
There is one orbital per site.
The Hubbard model takes the form 
 $\mathcal{H} = \mathcal{H}_0 + \mathcal{H}_1$, where $\mathcal{H}_0$ is the kinetic term 
 and $\mathcal{H}_1$ represents the onsite Hubbard interaction.
We consider the generic setting that has not been analyzed before, namely with the flat band of the noninteracting Hamiltonian being far away from the Fermi energy, as illustrated in Fig.\,\ref{fig:sch}a and shown in Fig.\,\ref{fig:dos}a. 

The lattice can be divided into two groups of sublattices (denoted by blue and yellow dots in Fig.\,\ref{fig:sch}d),
which contain different numbers of sites per unit cell.
The flat-band formation can be seen by 
considering only the nearest-neighbor hopping between the blue and yellow sites,
reflecting a 
destructive interference of the electronic wavefunction on the lattice
\cite{Ming2022} (see Supplementary Note\,2). 
The flat band overlaps with the wide bands.

\noindent
{\bf Molecular orbitals, effective extended Hubbard model and the solution method}

\noindent
A flat band that is topologically nontrivial
cannot by itself be represented by exponentially localized 
symmetry-preserving (Kramers-doublet) Wannier orbitals \cite{Cano2021}. Such a Wannierization only becomes possible when other bands are considered along with the flat band. 
In addition, a 
flat band coming from destructive interference comprises states from multiple (inequivalent) atomic sites. If one Wannier orbital is to primarily capture this flat band, this Wannier orbital (and, by extension, the others accompanying it) must involve multiple atomic orbitals. In other words, in this case, the Wannier orbitals are necessarily molecular orbitals. We again stress that the mapping we 
use is exact.

In our case, we can restrict to three
bands (see Supplementary Note\,3). 
 We find the centers of the three 
 localized Kramers-doublet 
 Wannier orbitals to be located near the geometric center of the unit cell, which forms a triangular lattice (see Fig.\,\ref{fig:sch}e). 
 Importantly,
 one Wannier orbital primarily captures
 the flat band \cite{Hu-flat-wide22.1};
 it is the most localized and 
 is denoted
 as the $d$ orbital. 
 The other two Wannier orbitals 
 are dominated by
 the wide bands; they
 decay much more slowly
 and are marked 
 as $c$ orbitals.
 The large
 difference in
 the width of the flat band ($D_{\rm flat}$) and 
 the wide bands ($D_{\rm wide}$) opens up a range of 
 interactions that are in between.
 In this range, the electron correlations are
 strongly orbital-selective 
and the system affords a Kondo/Anderson-lattice model description
\cite{Hu-flat-wide22.1}.

We project the Hubbard model of the original lattice to the Wannier basis. This leads to the effective model expressed in terms of the 
$d$ and $c$ Wannier orbitals with $H_{\mathrm{eff}}=H_{0}+H_{\rm int}$. The kinetic term is specified in the Methods.
For the interaction terms, we 
keep the most dominant interactions on the Wannier basis. They include the onsite Hubbard interaction 
among the $d$ electrons and 
the 
density-density
interactions 
between the $d$ and $c$ electrons:
\begin{equation}\label{eq:hint}
\begin{aligned}
    H_{\rm int} & = H_{d} + H_{F} \\
    & = \sum_{i} \frac{u}{2} \left(n_{i\up}^{d} + n_{i\dn}^d -1 \right)^2 + \sum_{i,\alpha} F_{\alpha} \, n_{i}^{d} \, n_{i}^{c,\alpha}
\end{aligned}
\end{equation}
where $n_{i,\sigma}^{a}=a^{\dagger}_{i\sigma} a_{i\sigma}$, with $a=d,c_{\alpha}$,  $\alpha=1,2$, and $n_{i}^{a} = \sum_{\sigma} n_{\sigma}^{a}$. 
The onsite Hubbard interaction on the $d$-orbital,
$u$, is the most dominant one, 
given the much more localized nature of this Wanner orbital.
The
density-density interactions between the $d$ and $c$ 
electrons, $F_{\alpha}$,
are 
weaker but 
also 
sizable:
$F_1/u\approx 0.3$ and $F_2/u\approx 0.25$. 
These effective interaction parameters are determined by those of the original Hubbard model 
(see Supplementary Note\,3).
The interactions among the $c$ electrons are 
relatively small compared to their bandwidths and, accordingly, will be unimportant.

To take into account the effect of the interactions, we use the U(1) slave spin (SS) method~\cite{Rong.12}. 
Given that only the onsite interaction of the $d$ orbital is important, we need to introduce a SS representation 
for the $d$ orbital only: $d_{i\sigma}^{\dagger}=o^{\dagger}_{i\sigma}f_{i\sigma}^{\dagger}$, where the auxiliary bosonic and fermionic operators, $o^{\dagger}$ and $f_{\sigma}^{\dagger}$, carry the charge 
and spin degrees of freedom, respectively.
We treat the SS formulation at the saddle-point level and 
 self-consistently solve the corresponding Hamiltonians for the SS and the auxiliary fermion parts.
The SS method is also used to obtain the contributions to the single-electron excitations 
from the (interaction-driven) 
incoherent part of the spectrum.
The details are found in the Methods and Supplementary Note\,6.

\noindent
{\bf Emergent flat band at the Fermi energy}  

\noindent
We are now in position to discuss the effect of interactions on the single-electron excitations. Consider first the density of states (DOS). In the noninteracting case, as shown in Fig.\,\ref{fig:dos}a, the $d$ electron  DOS (the red curve) has a 
sharp peak compared with the background (purple color) $c$ electron component. This reflects 
the $d$ electrons as primarily describing the flat band.
As can be seen, the peak is located far away from the  Fermi level, which also reflects its origin from the noninteracting flat band (see Supplementary Note\,1). 

Importantly, under the influence of electron correlations, a new flat band emerges.
This is demonstrated in Fig.\,\ref{fig:dos}b with $u=1.6$.  The emergent flat band is pinned to the Fermi energy, as captured by
the coherent peak (the red solid lines) in the DOS.
The background 
DOS associated with the conduction $c$ electrons is largely unchanged from its noninteracting counterpart. 
Varying the interaction 
strongly influences the 
spectral weight of the emergent flat band 
(see Supplementary
Note\,7):
This part of the spectral weight
is reduced as 
the interaction 
increases (comparing 
Fig.\,\ref{fig:dos}b
and Supplementary Fig.\,\ref{fig:dos_mott}a);
the reduced spectral weight is transferred to the incoherent part (the red dashed lines).
This form persists until the weight of the coherent peak is completely lost and the system goes through an orbital-selective Mott transition. When that happens, the incoherent parts of the single-electron excitations develop into the full-fledged lower and upper Hubbard bands (see Supplementary
Note\,7 and Fig.\,\ref{fig:dos_mott}b).

\noindent
{\bf Orbital-selective Mott correlations}

\begin{figure}[h]
\centering
\includegraphics[width=1\columnwidth]{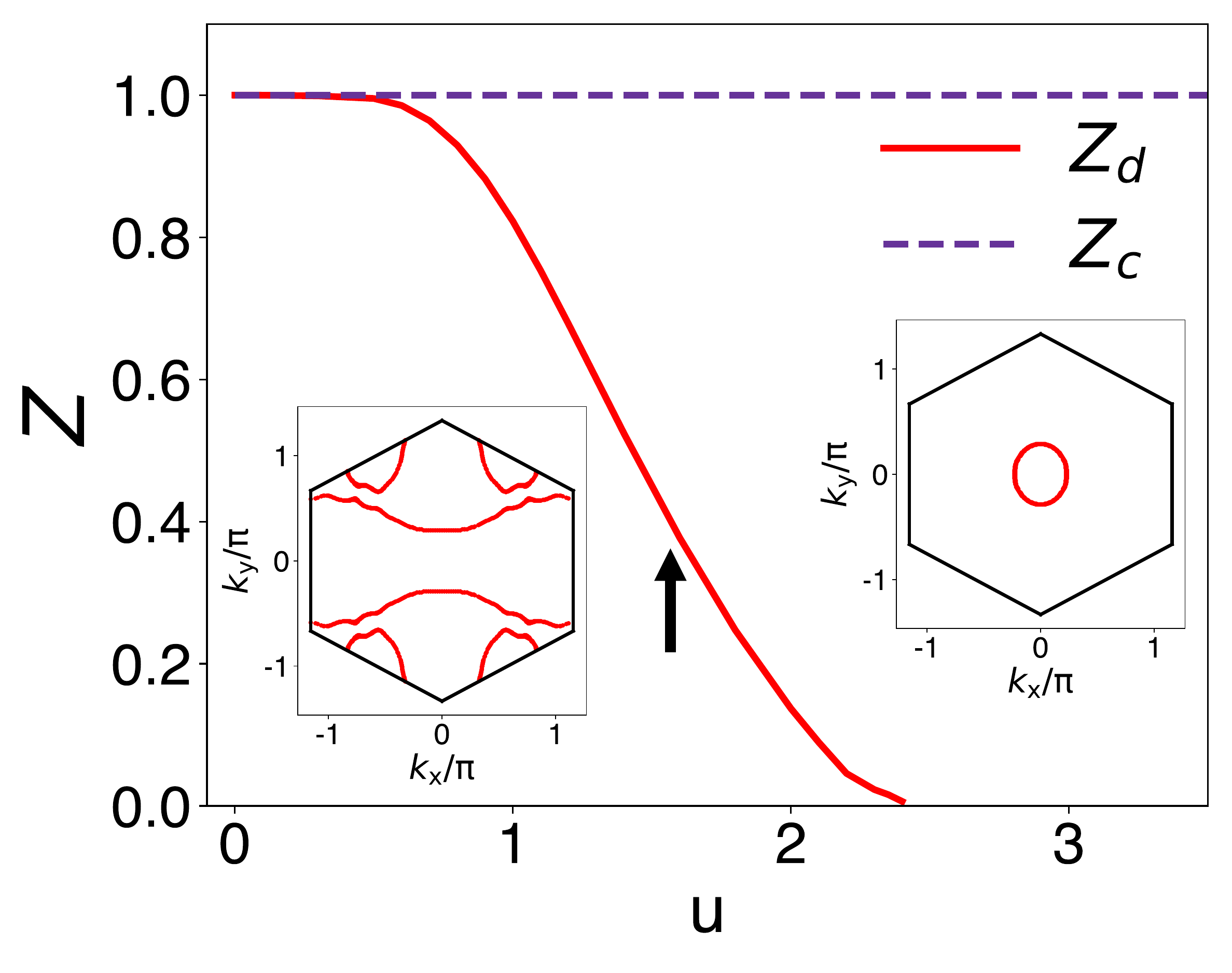}
\caption{
\textbf{Orbital-selective correlations.}
The quasiparticle weights of the $d$ and $c$ electrons,
$Z_d$ and $Z_c$, as a function of the effective interaction
$u$. The Left (right) inset plots 
the Fermi surface for the values of interactions 
right below (above) $u_c=2.4$.
}
\label{fig:qp}
\end{figure}

\noindent
To expound the origin of the emergent flat band, we  
further analyze the orbital-selective Mott correlations in the regime of interactions of our interest, viz. $D_{\rm flat} < u < D_{\rm wide}$. 
As shown in Fig.\,\ref{fig:qp},
the metal-to-insulator transition of the $d$ electrons occurs at $u_c=2.4$. 
We reiterate that the range of the interactions being considered here is weaker than the width of the wide bands associated with the $c$ electrons (see Supplementary Fig.\,\ref{fig:dos_c}). Thus, the $c$ electrons are fully itinerant. This justifies the neglect of 
the interactions among the $c$ electrons, as we have done,
so that the quasiparticle weight of the $c$-electrons
remains to be $1$ as seen in Fig.\,\ref{fig:qp}.
The existence of an orbital-selective Mott
transition is further illustrated 
in the nature of the Fermi surface. As shown in Fig.\,\ref{fig:qp} (insets),
the Fermi surface undergoes a dramatic change across the transition. This change
 parallels
 the 
 electron localization-delocalization (Kondo destruction) physics of 
 heavy fermion systems \cite{Si2001,Colemanetal,senthil2004a,paschen2004,
shishido2005}.
 The phase with the $d$-electrons being itinerant corresponds to the Kondo-screened phase, in which the local moment is converted into (fragile, or heavy) electronic excitations that  hybridize with the conduction electrons to form the quasiparticles. By contrast, the  orbital-selective Mott phase (OSMP) is analogous to the Kondo-destroyed phase of the heavy fermion systems, in which the Fermi surface is formed entirely from 
 the conduction electrons. 
 The fact that our noninteracting flat band is, to begin with, far away from the Fermi energy makes the parallel with the heavy fermion systems especially clear. 
 Our analysis of this 
 flat-band system provides 
 a realization of the Kondo physics in 
 a one-band Hubbard model that
 physically describes a $d$-electron-based system.

More specifically, the value of the 
interaction illustrated 
in Fig.\,\ref{fig:dos}b is marked by an arrow in Fig.\,\ref{fig:qp}. The differentiation between the quasiparticle weights of the $d$ and $c$ electrons at this interaction 
characterizes the
orbital-selective
nature of the electron correlations. 
 This is reflected in the $d$-electron spectral weight: as seen in Fig.\,\ref{fig:dos}b,
 the incoherent peaks (the red dashed lines) are well formed, which corresponds to the precursor of the lower and upper Hubbard bands of the OSMP (see Supplementary Fig.\,\ref{fig:dos_mott}b). 
 The coherent spectral weight, i.e. the central peak (the red solid line of Fig.\,\ref{fig:dos}b) is thus described in terms of the Kondo resonance of a Kondo lattice model, in which the local moments 
 correspond to the effective spin degrees of freedom associated with the lower and upper Hubbard bands.
 This description makes precise the notion that the flat band at the Fermi energy is emergent, driven by the orbital-selective Mott correlations.

\noindent
{\bf Topological Kondo semimetal} 

\begin{figure}[h]
\centering
\includegraphics[width=1\columnwidth]{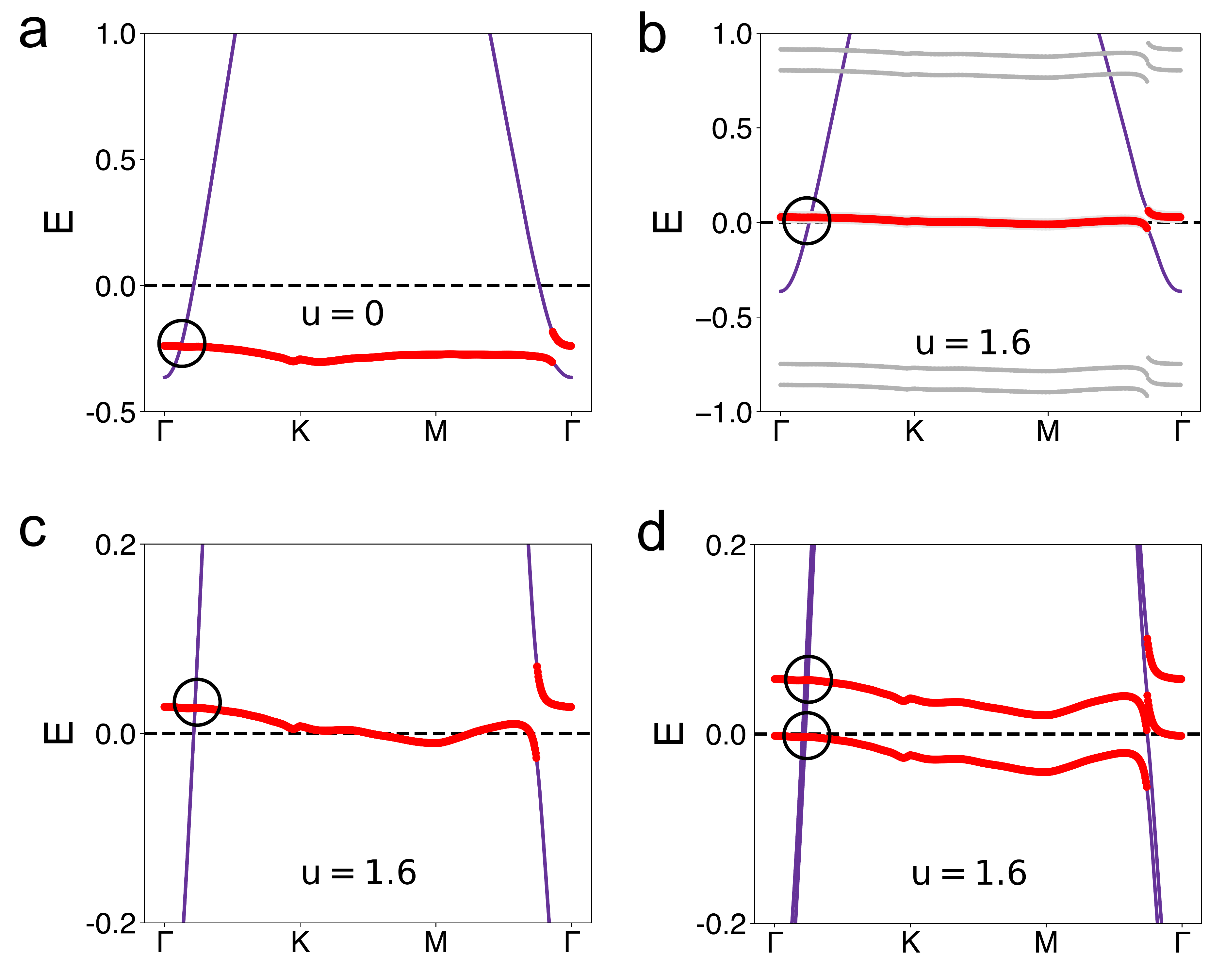}
\caption{\textbf{Topological Kondo semimetal.}
{\bf a,} The noninteracting band structure. 
{\bf b,} The dispersion of the single-electron 
excitations at 
$u=1.6$. The red solid curve denotes the emergent flat band close to the Fermi energy. The grey lines mark the incoherent single-electron excitations. 
{\bf c,} The zoomed-in view of the emergent flat band. 
{\bf d,} The band structure at $u=1.6$ with a Zeeman splitting $m_z=0.03$. 
}
\label{fig:disp}
\end{figure}

\noindent
The energy dispersion of the electronic states is shown in  Fig.\,\ref{fig:disp}.
From the dispersion of the interacting ($u=1.6$) case,
we again see that a Fermi-energy-bound flat band emerges 
in the interacting case. 

We are then in position to analyze the symmetry constraints\cite{Bradlyn2017,Cano2018,Po2017,Watanabe2017,Cano2021}. In the noninteracting limit, the three Wannier orbitals have different $M_x$ eigenvalues. The flat $d$ orbital has the $M_x$ eigenvalue $+1$, while the two $c$ orbitals have the $M_x$ eigenvalues of
$-1$ and $+1$ respectively~\cite{Hu-flat-wide22.1}. In the presence of time-reversal symmetry, along the $\Gamma-K$ line, the flat band from the $d$ orbital has a symmetry-protected Dirac crossing with the $c$ orbital of the opposite mirror eigenvalue.
This same symmetry constraint also applies to the Kondo-driven flat band. The Dirac node for
the emergent flat band
is shown in Fig.\,\ref{fig:disp}b.
A node from the flat band close to the Fermi energy allows a high tunability.
As shown in Fig.~\ref{fig:disp}d, 
a relatively
small Zeeman coupling (illustrated here with $m=0.03$, which is small compared to the width of the emergent flat band of $\sim 0.1$) 
causes a substantial separation of the nodes.
These nodes now have two-fold degeneracy.

The orbital-selective Mott correlations are caused by local correlations. While we have provided a case study
 of how such correlations give rise to emergent flat bands in a particular 2D model,
a similar conclusion is expected in general cases, including for models 
in three dimensions (3D).
There is an important distinction though.
In 3D, 
topological nodes develop in the presence of SOC 
under symmetry constraints \cite{Arm18.1,Cano2021}. For noncentrosymmetric systems, or for centrosymmetric systems with the breaking of
time-reversal symmetry, we can then expect the emergent flat bands to feature Weyl nodes leading to a Weyl Kondo semimetal.

To further expound on the generality of our theoretical results, we note that the 
metallic regime with strong orbital-selective correlations can be viewed through the Kondo analogy.
From this perspective, the emergent flat band describes low energy coherent electronic excitations associated with the Kondo-driven composite fermions. Because low energy electronic excitations are 
always Fermi-energy bound, and also based on the well-established understanding that Kondo-driven composite fermions occur in the immediate vicinity of the Fermi energy, the emergent flat bands that develop through our proposed mechanism must be pinned near the Fermi energy. This represents 
a general principle. To explicate on this generality, we have mapped out a phase diagram to show that the proposed mechanism operates over an extended region in the $u$-$\epsilon_d^0$ parameter space (region ``II" of the phase diagram given in Supplementary Note\,8 and Fig.\,\ref{fig:zall}). Furthermore, we have carried out related calculations 
in a more general setting and find a similar development of an emergent flat band when the noninteracting flat band is located substantially away from the Fermi energy; the details of this analysis appear in Supplementary Note\,9 and Fig.\,\ref{fig:kagome}.

\noindent
{\bf Design principle for Weyl Kondo semimetals in physical $d$-electron systems}

\noindent
Weyl Kondo semimetals have so far been explored in 
$f$-electron-based materials,
Ce$_3$Bi$_4$Pd$_3$ \cite{Dzsaber2017,Dzs-giant21.1} and several newly proposed Ce-, Pr- and U-based compounds \cite{Chen-Natphys22,Hu2021}. The present work leads us to propose a design principle for realizing Weyl Kondo semimetals in a new setting.
Importantly,
our theoretical results are expected to be robust against the effect of 
the residual interactions among the quasiparticles.
This is so because the origin of the topological nodes lies in the symmetry constraints,
which have recently been shown to operate 
on the eigenvectors of the matrix associated with the {\it exact} single-electron Green's function of an interacting system \cite{Hu2021}.

Accordingly, our proof-of-principle demonstration enables us to advance a new materials design procedure for Weyl Kondo semimetals in 
the setting of $d$-electron-based systems.
The procedure would start from 
3D lattices that can host 
flat bands from quantum interference. Examples include the pyrochlore lattice~\cite{Bergman2008}, the perovskite lattices~\cite{Weeks2010}, and other 3D versions of the bipartite crystalline lattices~\cite{Regnault2022}. We seek materials with
$d$-elements,
and utilize orbital-selective correlations to drive 
interacting flat bands that are Fermi-energy-bound. Symmetry constraints can then lead to either Dirac or (with the breaking of inversion or time-reversal symmetry) Weyl nodes in these emergent flat bands.
The latter case corresponds to a Weyl Kondo semimetal.

The procedure for this materials identification approach goes beyond that 
for Weyl Kondo semimetals in $f$-electron-based systems \cite{Chen-Natphys22}. 
In addition to the requirement for
both correlations and crystalline symmetry constraints, it also involves 
the crystal lattice conditions for the formation of 
flat bands in the bare dispersion. We reiterate that the noninteracting flat bands are not required to be near the Fermi energy. This is an important feature in the proposed materials design principle, given that the noninteracting flat bands in relevant materials are
generically 
away from their Fermi energy.

\noindent
{\bf Implications for fractional Chern insulators}

\noindent
Fractional Chern insulators, with a fractional quantum Hall effect and the associated fractional charge in a lattice setting, have been proposed in correlated models 
with an appropriate (such as $1/3$)
filling ratio of a flat band when the latter crosses the Fermi energy~\cite{Sheng2011,Titus2011,Regnault2011}. Experimental evidence has recently been identified in twisted bilayer graphene \cite{Xie-fci2021}, in which the moir\'{e} bands are located near the Fermi energy, in a small external magnetic field.
Our results on the Fermi-energy-bound emergent flat bands raise the possibility of 
another potential platform to realize the fractional Chern insulators, namely in
$d$-electron-based 2D systems.
Indeed, when 
a spin-orbit coupling is included 
in the 2D model, the Dirac node is
gapped leading to 
flat $Z_2$ topological bands
(see Supplementary Note\,4).
The residual interactions (which develop beyond the saddle-point analysis 
in the slave-spin approach that we have carried out) could be ferromagnetic
(see Supplementary Note\,5). In that case, the flat band can develop a nonzero Chern number and can be analyzed for a lattice realization of fractional quantum Hall effect \cite{Wang2021_PRL}.
Indeed,
the combination of the flatness of the associated bands (see Supplementary Fig.\,\ref{fig:bd_soc}) and the 
aforementioned 
residual interactions
among the heavy quasiparticles represents a condition that is
similar to what happens in the
moir\'{e} systems \cite{Xie-fci2021};
however, 
the $Z_2$ nature of the flat bands makes them distinct and rare \cite{Devakul-TBTMD2021}.
Accordingly, with appropriate fillings, our results suggest that the corresponding $d$-electron-based 2D materials provide a new setting for realizing
a fractional Chern insulator.
The naturalness of the emergent flat band crossing the Fermi energy makes our proposal robust.
Thus, this
represents a promising new direction for
a systematic examination.

\noindent
{\large \bf Discussion}

\noindent
Our work opens a new bridge 
between topological 
flat bands
and 
correlation physics. The interaction effect tends to localize the molecular orbital that has the most overlap with the flat band. As a result, these molecular orbitals play the role of local moments, by analogy with
the local spins of Kondo systems.
Our work provides a 
rare non-perturbative way
to
address the interplay between correlations and topology effects in 
such flat band systems 
and a variety of correlated 
materials \cite{CBCSP-NRM2023.x}.
As such, it
promises to elucidate the physics of 
correlated kagome transition-metal compounds~\cite{Ye2018,Yao18.1x,Ye2021.x,Ekahana2021.x} and related materials.
We also expect that our analysis 
will inspire new understandings of the correlation effects in
moir\'e structures, 
which are increasingly being viewed from a Kondo perspective~\cite{Zhao2022.2,Ram2021,Song2022,Guerci2022.x,Chou2022}, as well as in other flat band systems~\cite{Addison2022}.
Our work has also allowed us to advance a new materials design principle to identify 
Weyl Kondo 
semimetals in the new setting of $d$-electron-based systems. We expect the interconnections that our work reveals among 
seemingly disparate systems to inspire new realizations and understandings of correlated topological effects in a wide variety of quantum materials and structures.
Finally, we note that our theoretical result for the emergent flat band is now supported by experiment:
%
%{\bf Note added}: 
%%Since the initial submission,
%%of the manuscript, 
%our work has motivated a new experiment: 
In a frustrated-lattice material, an emergent flat band has been 
%experimentally 
observed by angle resolved photoemission spectroscopy at the Fermi energy, even though the {\it ab initio} noninteracting band structure predicts a flat band that is
considerably away from the Fermi energy~\cite{Huang2023np}.
\vskip 1 cm

%%%%%%%%%%%%%%
\noindent{\bf\large Methods}
\\
%\\
\noindent
{\bf Hubbard model on the clover lattice}

\noindent
The clover lattice has been discovered in real materials
such as the van der Waals 
system
$\mathrm{Fe_{5}GeTe_{2}}$~\cite{Ming2022,May2019}. As shown in Fig.~\ref{fig:sch}d, it contains five sites in each unit cell,
which are reclassified into two groups as marked by the yellow and blue colors. 
For an illustrative purpose,
we restrict our model to  have only a $d_{z^2}$ orbital on each site. The case with other $d$-orbitals, such as $d_{xz}/d_{yz}$,
have a similar realization of the 
geometry-induced flat bands~\cite{Ming2022}. 
We consider the Hubbard model written as $\mathcal{H} = \mathcal{H}_0 + \mathcal{H}_1$, where $\mathcal{H}_0$ is the kinetic term that connects the two different groups of sublattices and $\mathcal{H}_1$ represents the onsite one-orbital Hubbard interaction.  We label the orbitals based on the group of sublattices to be $A/B$ and $C/D/E$ respectively. For each site, we consider the onsite Hubbard interaction,
\begin{equation}
    \mathcal{H}_1 = U\sum_{i,\alpha} n_{i,\up}^{\eta_{\alpha}} n_{i,\dn}^{\eta_{\alpha}} \, ,
\end{equation}
where $\eta_{\alpha}$ ($\alpha=1\sim 5$) goes through all the five orbitals in each unit cell. The kinetic Hamiltonian is written as
\begin{equation}
    \mathcal{H}_0 = \sum_{ij,\alpha\beta,\sigma} t \eta_{i\alpha\sigma}^{\dagger} \eta_{j\beta\sigma} -\mu_0 \sum_{i\alpha\sigma} \eta^{\dagger}_{i\alpha\sigma} \eta_{i\alpha\sigma} + \sum_{i,\sigma,\alpha\in \{ C,D,E\}} m\eta_{i\alpha\sigma}^{\dagger}\eta_{i\alpha\sigma} + \sum_{i\sigma\alpha\in \{D,E\}} \gamma \eta^{\dagger}_{i\alpha\sigma}\eta_{i\alpha\sigma} \, .
\end{equation}
Here $t$ denotes the
nearest-neighbor hopping between the two sites that are connected by the solid lines shown in Fig.~\ref{fig:sch}d, and $\mu_0$ is the chemical potential. 
In addition,
$m$ denotes the energy splitting between the two 
groups of sublattices, which 
have a different local environment and thus generically have different energy levels.
Finally, $\gamma$ represents an additional energy splitting between $C$ and $D/E$, which breaks the $C_{3}$ rotational symmetry.
As mentioned earlier, we work with the case that 
breaks $C_3$ symmetry to simplify the symmetry characterization and, thus, the Wannier construction.
It is possible that this $C_3$ symmetry breaking spontaneously appears as a result of interactions that drive a nematic order, although, for our illustrative purpose, we do not pursue this route specifically.
A detailed analysis of the dispersion is shown in Supplementary Note\,1. 

\noindent
{\bf Effective extended Hubbard model}

\noindent
We project the Hubbard model of the original lattice to the
Wannier basis. This leads to the effective model expressed in terms of the $d$ and $c$ Wannier orbitals with $H_{\mathrm{eff}}=H_{0}+H_{\rm int}$. The kinetic term takes the following form:
\begin{equation}
\label{eq:hk}
\begin{aligned}
     H_{0} & = H_{d} + H_{c} + H_{V}\\
     & = \sum_{ij, \sigma} t_{ij} \left( d_{i\sigma}^{\dagger} d_{j\sigma} +h.c. \right) - \sum_{i} \mu d_{i\sigma}^{\dagger} d_{i\sigma} \\
     & + \sum_{ij,\alpha\beta,\sigma} t_{ij}^{\alpha\beta} \left(c_{i\alpha\sigma}^{\dagger} c_{j\beta\sigma} + h.c. \right) + \sum_{i\alpha\sigma} (\Delta_{\alpha} -\mu) c_{i\alpha\sigma}^{\dagger} c_{i\alpha\sigma} \\
     & + \sum_{ij,\alpha\sigma} V_{ij}^{\alpha} \left(d_{i\sigma}^{\dagger} c_{j\alpha\sigma}  +h.c. \right) \, .
\end{aligned}
\end{equation}
Here, $d^{\dagger}_{i\sigma}$ ($c^{\dagger}_{i\alpha\sigma}$ ) creates a heavy (light) electron at the position $i$ with spin $\sigma$ (and orbital $\alpha$). In addition, $t_{ij}$ ($t_{ij}^{\alpha\beta}$) denotes the hopping parameter between the $d$ ($c$) electrons at the positions $i$ and $j$ (orbitals $\alpha$ and $\beta$ ). Moreover, $V^{\alpha}$ represents the 
hybridization between the light orbital $\alpha$
and the heavy orbital.
Finally, 
$\Delta_{\alpha}$ describes the difference in the energy levels between the $d$ orbital and $c$ orbitals, and $\mu$ specifies the chemical potential.
We will focus on the case with the $d$-electron level being deep below 
 the Fermi energy
and will show that the interaction effect 
creates a heavy band 
near the Fermi energy. 
The noninteracting dispersion 
is shown in Fig.\,\ref{fig:disp}a, where there is a Dirac crossing between the flat band and lower wide band located deep below the Fermi energy; the crossing is 
protected by the $M_{x}$ lattice symmetry~\cite{Cano2021}.  

\noindent
{\bf Slave spin method and self-consistent equations}

\noindent
We describe 
the 
U(1) slave spin approach~\cite{Rong.12}.
Because the bandwidth of the heavy orbital is much smaller than those of the light orbitals, the interaction effect is most pronounced on the $d$ orbital. We therefore only introduce the SS representation on the $d$ orbital: $d_{i\sigma}^{\dagger} = o^{\dagger}_{i\sigma} f_{i\sigma}^{\dagger}$.
The auxiliary bosonic field $o^{\dagger}_{\sigma} = P^{+}S^{+}_{\sigma}P^{-}$ is represented by the spin operator accompanied with the projection operators $P^{\pm}_{i\sigma} = \frac{1}{\sqrt{1/2 \pm S^{z}_{i\sigma}}}$ suitable for a system that is away from half filling. We treat the SS formulation at the saddle-point level by fully decoupling the SS and auxiliary fermion operators.
This leads to the decoupled Hamiltonian: 
\begin{equation}
\label{eq:self}
\begin{aligned}
     H^{f} &= \sum_{\kbf, \sigma} \langle O_{\sigma}\rangle^2 \epsilon_{\kbf}^{f}  f_{\kbf\sigma}^{\dagger}f_{\kbf\sigma}  + \sum_{\kbf\sigma} \left( -\mu -\lambda_{\sigma} + \lambda_{\sigma}^{0}  \right) f_{\kbf\sigma}^{\dagger}f_{\kbf\sigma} \\
     & +\sum_{\kbf,\alpha\sigma}  \left( V_{\kbf}^{\alpha} \langle O_{\sigma}^{\dagger} \rangle f_{\kbf\sigma}^{\dagger}c_{\kbf\alpha\sigma} + h.c. \right) +H_{c}  \\
      H^{S} &= \sum_{\sigma}\left[ \Tilde{\epsilon}_{f} \left( \langle O_{\sigma}\rangle O_{\sigma}^{\dagger} + h.c. \right) +\sum_{\alpha} \left( \Tilde{V}_{\alpha}O_{\sigma}^{\dagger} + h.c. \right) \right] \\
      &+\sum_{\sigma} \lambda_{\sigma} \left(S^{z}_{\sigma} + \frac{1}{2} \right) + H_{\rm int}^{S} \, ,
\end{aligned}
\end{equation}
where 
\begin{equation}
\label{eq:intS}
    H_{\rm int}^{S} =  \frac{u}{2} (\sum_{\sigma} S^{z}_{\sigma} )^2 + \sum_{\alpha} F_{\alpha} \langle n^{\alpha}_c \rangle \sum_{\sigma}S^{z}_{\sigma} \, ,
\end{equation}
and $\epsilon_{f}(\kbf)$ and $V^{\alpha}_{\kbf}$ are the Fourier transforms of $t_{ij}$ and $V^{\alpha}_{ij}$, respectively,
$\Tilde{\epsilon}_{f} = \sum_{\kbf} \epsilon_{f}(\kbf) \langle f^{\dagger}_{\kbf} f_{\kbf} \rangle$, and $\Tilde{V}_{\alpha} = \sum_{\kbf} V_{\kbf}^{\alpha} \langle f_{\kbf}^{\dagger} c_{\kbf} \rangle$. In addition, $O_{\sigma} = \langle P^{+} \rangle S_{\sigma}^{+} \langle P^{-}\rangle$, $\lambda_{\sigma}^{0} = 2\left( \bar{\epsilon}_{f} + \sum_{\alpha} \bar{V}_{\alpha} \right) \eta_{\sigma}$ with $\bar{\epsilon}_{f} = \Tilde{\epsilon}_{f} \langle O_{\sigma} \rangle \langle O_{\sigma}^{\dagger}\rangle + c.c.$, $\bar{V} = \Tilde{V} \langle O_{\sigma}^{\dagger} \rangle + c.c.$ and $\eta_{\sigma} = \frac{1}{2} \frac{n^f_{\sigma} -1/2}{(1-n^f_{\sigma})n^f_{\sigma} }$. Finally, we 
introduce the Lagrangian multiplier $\lambda_{\sigma}$  to remove the unphysical Hilbert space (see Supplementary Note\,6). The pseudo-spin carries the U(1) charge degree of freedom;
 the quasiparticle weight associated with the coherent part near the Fermi level is described by $Z = \langle O_{\sigma} \rangle \langle O^{\dagger}_{\sigma} \rangle$. 
An (orbital-selective) Mott localization transition happens when some quasiparticle weight $Z$ goes to zero.

In addition to the coherent quasiparticle peak, the SS method 
also calculates the contributions from the incoherent excitations. The Green's function of the $d$ electron $G_{d}$ is obtained by the convolution of the SS and the auxiliary fermions, with $G(\kbf, i\omega_{n}) = \sum_{i\Omega_m}G_{S}(i\Omega_m) G_{f}(\kbf, i\omega_{n} - i\Omega_{m})$, where $G_{S}(\tau) = \langle {\rm T}_{\tau} o_{\sigma} (\tau) o^{\dagger}_{\sigma}(0)\rangle$ and $G_{f}(\tau) =-\langle {\rm T}_{\tau} f_{\sigma}(\tau) f_{\sigma}^{\dagger}(0) \rangle$. 
The spectral function is then obtained from $A(\kbf,\omega) = \frac{-1}{\pi} \Im G_{d}^{R}(\kbf,\omega)$.

\vskip 1 cm
\noindent{\bf\large Data availability}
\\
The data that support the findings of this study are either presented in the manuscript or available at \href{https://doi.org/10.5281/zenodo.11247849}{https://doi.org/10.5281/zenodo.11247849}.

\vskip 1 cm
\noindent{\bf\large Code availability}
\\
The computer codes that were used to generate the data that support the findings of this study are available from the corresponding authors upon request.

% %%%%%%%%%%
\bibliographystyle{naturemagallauthors}
% %%%%%%%%%%%%%%%
\bibliography{osmp}

\clearpage

\clearpage

\noindent{\bf Acknowledgment}
\\
We thank Gabriel Aeppli, Joseph Checkelsky, Han Wu, Yonglong Xie and Ming Yi for useful
discussions. Work at Rice has primarily been supported by the U.S. DOE, BES, under Award No.
DE-SC0018197 (model construction, L.C.,F.X.,S.S.), by the Air Force Office of Scientific Research under Grant No.
FA9550-21-1-0356 (orbital-selective Mott transition, L.C.,F.X.,S.S.,H.H.,Q.S.), by the Robert A. Welch Foundation Grant No. C-1411 (model calculation, L.C.), and 
by the Vannevar Bush Faculty Fellowship ONR-VB N00014-23-1-2870 (conceptualization, Q.S.).The
majority of the computational calculations have been performed on the Shared University Grid
at Rice funded by NSF under Grant EIA-0216467, a partnership between Rice University, Sun
Microsystems, and Sigma Solutions, Inc., the Big-Data Private-Cloud Research Cyberinfrastructure
MRI-award funded by NSF under Grant No. CNS-1338099, and the Extreme Science and
Engineering Discovery Environment (XSEDE) by NSF under Grant No. DMR170109. H.H. acknowledges
the partial support of the European Research Council (ERC) under the European
Union's
Horizon 2020 research and innovation program (Grant Agreement No.\ 101020833). Work in
Vienna was supported by the Austrian Science Fund (project I 5868-N - FOR 5249 - QUAST)
and the ERC (Advanced Grant CorMeTop, No.\ 101055088). J.C. acknowledges the support of
the National Science Foundation under Grant No. DMR-1942447, support from the Alfred P.
Sloan Foundation through a Sloan Research Fellowship and the support of the Flatiron Institute,
a division of the Simons Foundation. J.C. and Q.S. acknowledge the hospitality of the Aspen Center for
Physics, which is supported by NSF grant No. PHY-2210452.

\vspace{0.2cm}
\noindent{\bf Author contributions}
\\
Q.S. conceived the research. L.C., F.X., S.S., H.H., J.C. and Q.S. carried out 
model studies.
L.C., S.P., J.C. and Q.S. contributed to
the development of 
the design principle for
correlated topological phases in $d$-electron systems. 
L.C. and Q.S. wrote the manuscript, with inputs from all authors.
%\\

\vspace{0.2cm}
\noindent{\bf Competing 
 interests}\\
The authors declare no competing 
 interests.
 %\\
 
 \vspace{0.2cm}
 \noindent{\bf Additional information}\\
Correspondence and requests for materials should be addressed to 
Q.S. (qmsi@rice.edu)
% \\

\clearpage

%%%%%%%%%%%%%
%%%%%%%%%%%%%%%%%%%%%%%%%%%%%%%%%%%%%%%

\clearpage

%%%%%%%%%%%%%
%%%%%%%%%%%%%%%%%%%%%%%%%%%%%%%%%%%%%%%

\clearpage

%%%%%%%%%%%%%
%%%%%%%%%%%%%%%%%%%%%%%%%%%%%%%%%%%%%%%

\clearpage

 %%%%%%%%%%%%%
%%%%%%%%%%%%%%%%%%

\newpage

\onecolumngrid 
\begin{center}
\textbf{\large Supplementary information for: Emergent flat band and topological Kondo semimetal driven by orbital-selective correlations}
\end{center}

\setcounter{secnumdepth}{2} % default value for 'report' class is "2"
\setcounter{equation}{0}
\setcounter{figure}{0}
\setcounter{table}{0}
\renewcommand{\theequation}{S\arabic{equation}}
\renewcommand{\thefigure}{S\arabic{figure}}

% %%%%%%%%%%
% \bibliographystylesupp{naturemagallauthors}
% %%%%%%%%%%%%%%%
% \bibliographysupp{osmp}
% \renewcommand{\bibnumfmt}[1]{[S#1]}
% \renewcommand{\citenumfont}[1]{S#1}

\section*{Supplementary Note 1: Band structure of 
%in 
the original Hamiltonian}\label{Sec:bdorg}
% \subsection*{Band structure of in the original Hamiltonian}
% \section{Band structure of in the original Hamiltonian}\label{Sec:bdorg}

\begin{figure}[h]
\centering
\includegraphics[width=0.9\columnwidth]{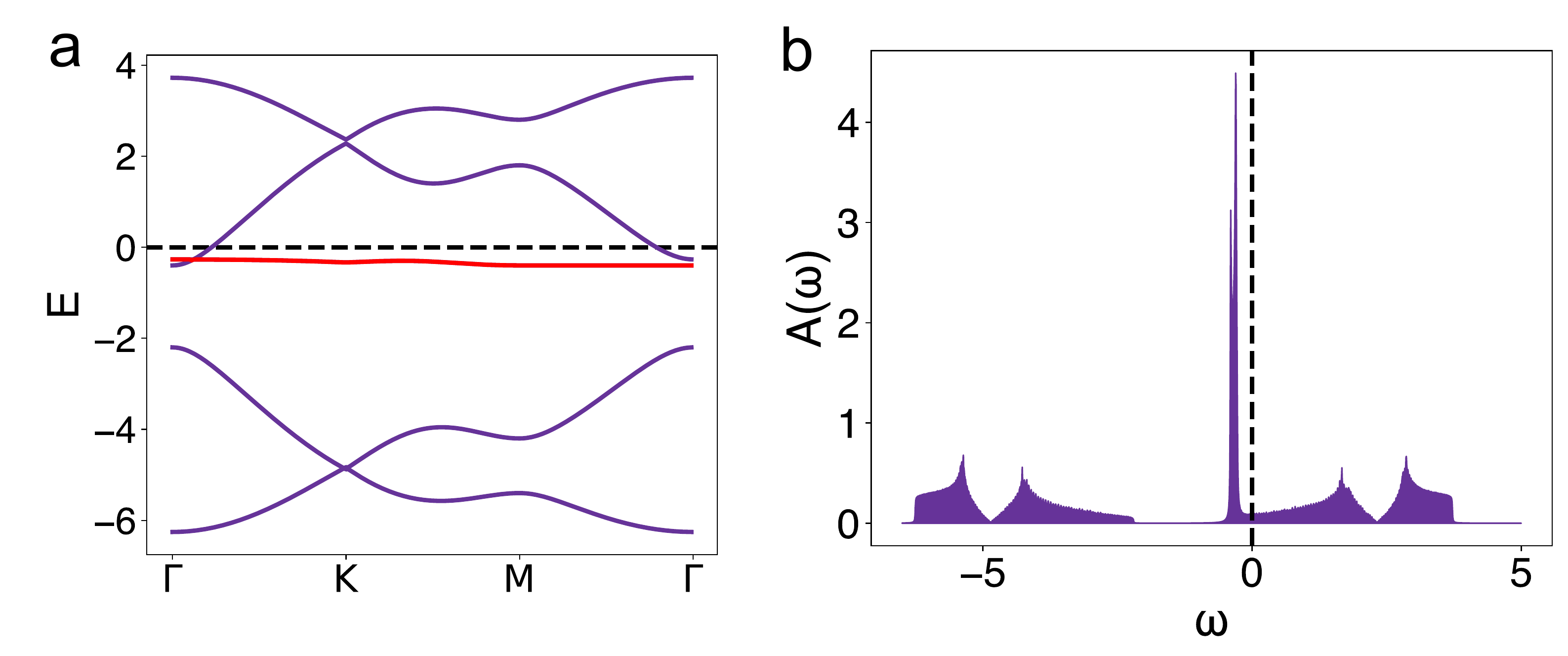}
\caption{ \textbf{The noninteracting electronic structure of the original Hubbard model.}
{\bf a,} The band structure in the original clover lattice, with $\mu=2.2$, $m=2$ and $\gamma=-0.2$.
{\bf b,} The corresponding density of states.
}
\label{fig:bd_org}
\end{figure}

In this section, we show the dispersion in the original lattice.  We fix the reference energy scale $t=2$. The noninteracting band structure is shown in Fig~\ref{fig:bd_org}a, in which the bands are clearly separated into the lower two bands and upper three bands. The spectral function is provided in Fig.~\ref{fig:bd_org}b, where the sharp peak denotes the position of the flat band. We focus on the upper three bands for our Wannier construction. 

 %%%%%%%%%%%%%
%%%%%%%%%%%%%%%%%%

\section*{Supplementary Note 2: Destructive interference and the origin of the flat band}\label{Sec:geometric-flat-band}
\begin{figure}[h]
\centering
\includegraphics[width=0.4\columnwidth]{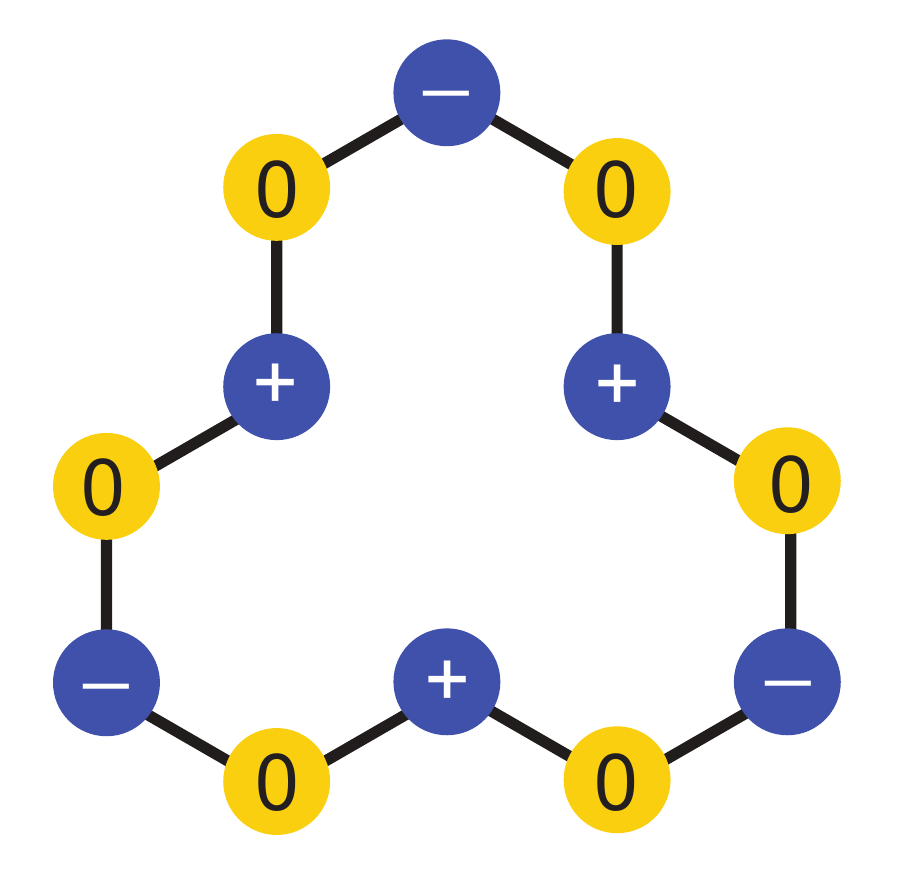}
\caption{ 
{\bf Wavefunction for the electronic states of the flat band.} Illustrated here is the real space representation of the electronic wavefunction for the flat band in the clover lattice.
}
\label{fig:fb_wf}
\end{figure}

% \subsection*{Destructive interference and the origin of the flat band}
% \section{
% Destructive interference and the origin of the flat band
% }
% \label{Sec:geometric-flat-band}
In this section, we discuss the origin of the flat band in the clover lattice. Consider the 
nearest-neighbor hoppings between the blue and yellow sublattices. The kinetic part of the Hamiltonian takes the form as
\begin{equation}
    \mathcal{H}_0(\kbf) = \begin{pmatrix}
    0_{2\times 2} & \mathcal{H}_{\kbf} \\
    \mathcal{H}_{\kbf} & 0_{3\times 3}
    \end{pmatrix} \, ,
\end{equation}
where $\mathcal{H}_{\kbf}$ is the hopping matrix between the two groups of sublattices. 
The localized wavefunction takes the following form~\cite{Ming2022}:
\begin{equation}
    \psi(k_x, k_y) = \begin{pmatrix}
    0 & 0 & e^{\frac{k_y}{3}} - e^{-\frac{2}{3}k_y} & e^{-\frac{k_x}{2\sqrt{3}} - \frac{k_y}{6}} -  e^{\frac{k_x}{\sqrt{3}} + \frac{k_y}{3}} & e^{\frac{k_x}{2\sqrt{3}} - \frac{k_y}{6}} - e^{-\frac{k_x}{\sqrt{3}} + \frac{k_y}{3}}
    \end{pmatrix} \,.
\end{equation}
Here, the amplitude is identically zero on the yellow sites because of the destructive interference effect. The real space wavefunction is shown in Fig.\,\ref{fig:fb_wf}. 

 %%%%%%%%%%%%%
%%%%%%%%%%%%%%%%%%

\section*{Supplementary Note 3: Effective Multi-orbital Hubbard Model}\label{Sec:Wannier}
\begin{figure}[h]
\centering
\includegraphics[width=0.9\columnwidth]{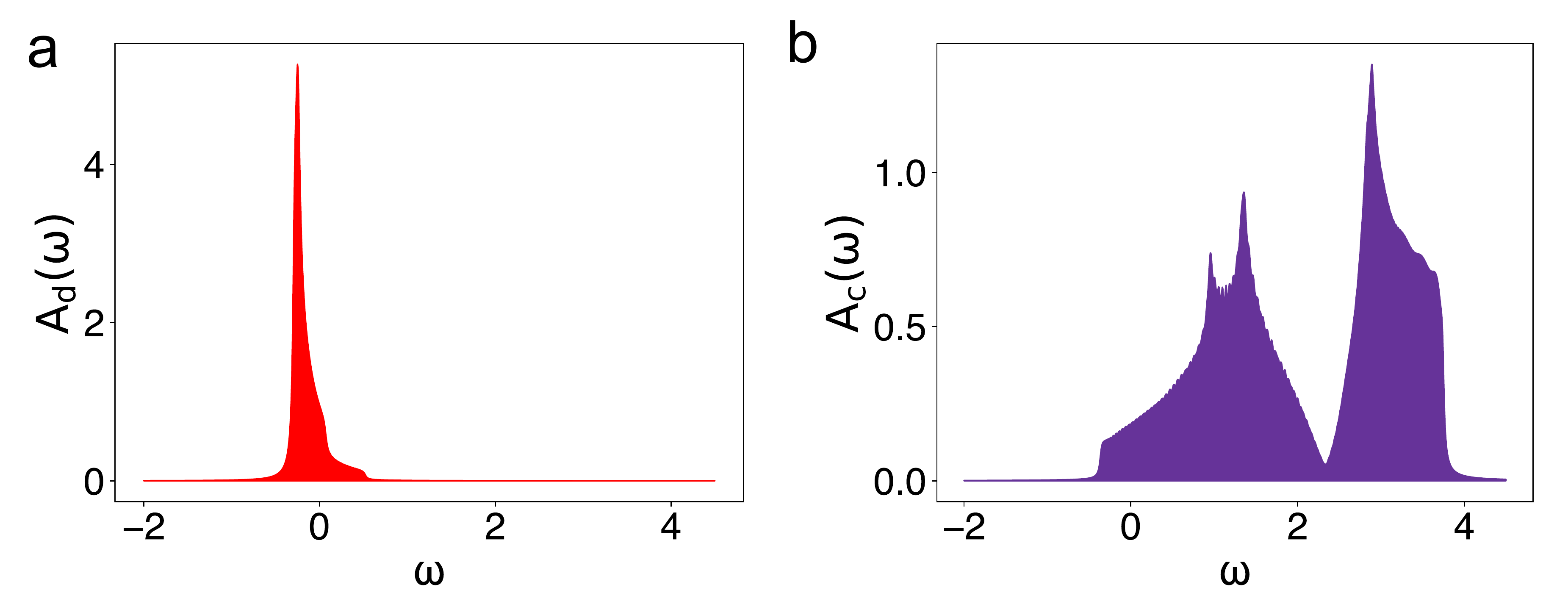}
\caption{{\bf The noninteracting density of states projected to the Wannier orbitals.}
{\bf a,b,} The 
DOS of the $d$ and $c$ electrons, respectively, in the absence of 
any hybridization between the two orbitals.
}
\label{fig:dos_c}
\end{figure}

% \subsection*{Effective Multi-orbital Hubbard Model}
% \section{Effective Multi-orbital Hubbard Model}
% \label{Sec:Wannier}
For the kinematic hoppings of the effective molecular orbitals, we refer to Table II in the SM of Ref.~\citenum{Hu-flat-wide22.1}. However, we focus on the case such that, in the noninteracting case, the flat band lies considerably below the Fermi energy.
In Fig.\,\ref{fig:dos_c}, we present the DOS for the $d$ and $c$ electrons in the absence of 
any hybridization between the two orbitals.
For the interaction part, the ratio between the Hubbard interaction of the effective $d$ electron and the one on the original lattice is $u/U=0.15$~\cite{Hu-flat-wide22.1}. For the same reason that the onsite interaction among the $c$-orbitals is unimportant, the Hund’s coupling does not play an important role here~\cite{Hu-flat-wide22.1}.

 %%%%%%%%%%%%%
%%%%%%%%%%%%%%%%%%

\section*{Supplementary Note 4: Analysis of the Effective Multi-orbital Hubbard Model with a Spin-Orbit Coupling}\label{Sec:soc}
% \subsection*{Analysis of the Effective Multi-orbital Hubbard Model with a Spin-Orbit Coupling}
% \section{Analysis of the Effective Multi-orbital Hubbard Model with a Spin-Orbit Coupling}\label{Sec:soc}
In this section, we discuss the influence of the spin-orbit coupling (SOC). For our purpose, it is adequate to consider a SOC that is small compared to the other 
energy scales. We can then analyze the effect of the SOC perturbatively, by using
the same Wannier orbitals described above.
We thus project the SOC of the original lattice onto the
Wannier orbital basis. We consider the SOC along 
the $z$ direction of the original clover lattice, which takes the form as
\begin{equation}
    \mathcal{H}_{\rm soc} = \sum_{i, \sigma =\pm } i t_{soc}\sigma \left( \eta_{i\sigma C }^{\dagger} \eta_{i\sigma D} +\eta_{i\sigma D }^{\dagger} \eta_{i\sigma E} + \eta_{i\sigma E }^{\dagger} \eta_{i\sigma C}\right) + h.c. \, .
\end{equation}
Here, $t_{soc}$ is the strength of the SOC, and $\sigma=\pm$ represents spin up and spin down, respectively. The hopping parameters on the basis of the three Wannier orbitals are shown in Supplementary Table.~\ref{tab:hop_soc_1} and Supplementary Table.~\ref{tab:hop_soc_2}.

\begin{table}[h]
    \centering
    \begin{tabular}{c|c |c | c| c  }
    \hline
      $(i,j)$ &  $(0,1)$ & $(-1,0)$ & $(1,0)$ & $(0,-1)$ \\
      \hline
      $t_{ia_1+ja_2}$ & $-0.04566 + i 0.01088$ & $-0.04566 - i 0.01088$ & $-0.04566 + i 0.01088$ &$-0.04566 - i 0.01088$ \\
      \hline
      &  $\pm (2,-2)$ & (0, -2) & (0, 2) & (2,0)  \\
      \hline
      & 0.04991 & $0.03701-i0.0031$ & $0.03701+i0.0031$ & $0.03701+i0.0031$ \\
      \hline
      &  $(-2,0)$ & (-1, 2) & (1,2) & (-2,1) \\
      \hline
      & $0.03701-i0.0031$ &$-0.03314-i0.00061$ &$-0.03314+i0.00061$ &$-0.03314+i0.00061$\\
      \hline
       &  $(2,1)$ & $\pm$(1, -1)  \\
      \hline
      & $-0.03314-i0.00061$ & $0.02032$\\
      \hline\hline
      $(i,j)$ &  $(1,0)$ & $(0,-1)$ & $(-1,2)$ & $(-2,1)$ \\
      \hline
      $V_{ia_1+ja_2}^{1}$& $0.13301+i0.01893$ &$-0.13301+i0.01893$ &$0.09264-i0.00299$ &$-0.09264-i0.00299$\\
      \hline
      &  $(1,-2)$ & $(2,-1)$ & $(0,-2)$ & $(2,0)$ \\
      \hline
       & $0.07903-i0.00749$ &$-0.07903-i0.00749$ &$-0.06361+i0.00909$ &$0.06361+i0.00909$\\
       \hline
      &  $(-2,3)$ & $(-3,2)$ & $(1,1)$ & $(-1,-1)$ \\
      \hline
       & $0.05494-i0.00431$ &$-0.05494-i0.00431$ &$-0.05082-i0.007$ &$0.05082-i0.007$\\
       \hline\hline
      $(i,j)$ &  $(1,-1)$ & $(-1,1)$ & $(0,1)$ & $(-1,0)$ \\
      \hline
      $V_{ia_1+ja_2}^{2}$& $-0.15053 $ &$-0.14577$ &$0.096-i0.00896$ &$0.096+i0.00896$\\
      \hline
      &  $(0,-1)$ & $(1,0)$ & $(0,0)$ & $(2,-2)$ \\
      \hline
      & $-0.09143 +i0.01766$ &$-0.09143 -i0.01766$ &$-0.07918$ &$0.07563$\\
      \hline
      &  $(1,1)$ & $(-1,-1)$ & $(-1,2)$ & $(-2,1)$ \\
      \hline
      & $-0.07542 +i0.00449$ &$-0.07542 -i0.00449$ &$0.07311+i0.00509$ &$0.07311-i0.00509$\\
      \hline\hline
    \end{tabular}
    \caption{{\bf Parameters for the noninteracting Hamiltonian in the Wannier basis -- part I.} Shown here are the
    hopping parameters for the $d$ orbitals and the hybridization between the $d$ and $c$ orbitals for spin up when $t_{\rm soc}=0.06$. The parameters in the spin-down case are the Hermitian conjugate of their spin-up counterparts. }
    \label{tab:hop_soc_1}
\end{table}
\begin{table}[h]
    \centering
    \begin{tabular}{c|c |c | c| c  }
        \hline
       $(i,j)$ &  $(0,-1)$ & $(0,1)$ & $(1,0)$ & $(-1,0)$ \\
      \hline
      $t^{11}_{ia_1+ja_2}$ &$-0.35388+i0.01913 $  & $-0.35388-i0.01913$& $-0.35388-i0.01913$& $-0.35388+i0.01913$\\
      \hline
       &  $\pm(1,-1)$ & $\pm (2,-2)$ & (-1,-1) & (1,1) \\
      \hline
       &$-0.15797 $  & $-0.08055$ & $-0.07438+i0.00105$ & $-0.07438-i0.00105$ \\
      \hline\hline
      $(i,j)$ &  $(0,-1)$ & $(0,1)$ & $(1,0)$ & $(-1,0)$ \\
      \hline
      $t^{22}_{ia_1+ja_2}$ &$0.27789-i0.00826 $  & $0.27789+i0.00826$& $0.27789+i0.00826$& $0.27789-i0.00826$\\
      \hline
       &  $(0,-2)$ & $(0,2)$ & $(-2,0)$ & $(2,0)$ \\
      \hline
       & $-0.07658+i0.0008$& $-0.07658-i0.0008$& $-0.07658+i0.0008$ & $-0.07658-i0.0008$\\
       \hline
       &  $(-1,-1)$ & $(1,1)$  \\
       \hline
       & $0.05848+i0.0014$ & $0.05848-i0.0014$\\
       \hline\hline
       $(i,j)$ &  $(0,-1)$ & $(1,0)$ & $(1,-2)$ & $(2,-1)$ \\
      \hline
      $t^{12}_{ia_1+ja_2}$ &$0.43550-i0.00395 $  & $-0.43550-i0.00395$& $0.1959-i0.00079$& $-0.1959-i0.00079$\\
      \hline
       &  $(0,1)$ & $(-1,0)$ & $(2,-3)$ & $(3,-2)$ \\
      \hline
       &$0.15083-i0.00138 $  & $-0.15083-i0.00138 $& $0.11537-i0.00017$& $-0.11537-i0.000179$\\
      \hline
      &  $(0,2)$ & $(-2,0)$ & $(-1, 2)$ & $(-2,1)$ \\
      \hline
       &$0.09257+i0.00133 $  & $-0.09257+i0.00133 $& $0.09107+i0.00516$& $-0.09107+i0.00516$\\
      \hline\hline
      & $\mu$ &  $\Delta_1$ & $\Delta_2$ \\
      \hline
      & $0.15$ & $1.81$ & $2.708$\\
      \hline
    \end{tabular}
    \caption{{\bf Parameters for the noninteracting Hamiltonian in the Wannier basis -- part II.} Shown here are the hopping parameters between the $c$ electrons 
    for spin up when $t_{\rm soc}=0.06$, the chemical potential $\mu$ and the effective crystal field differences $\Delta_{\alpha}$. The parameters for the spin-down case are the Hermitian conjugate of their spin-up counterparts.}
    \label{tab:hop_soc_2}
\end{table}

Because the value of the SOC is small,
we use the same saddle-point 
parameters as we obtained in the case without 
the SOC. The renormalization factors $Z$ for the different orbitals will then renormalize the SOC accordingly.

\begin{figure}[h]
\centering
\includegraphics[width=0.9\columnwidth]{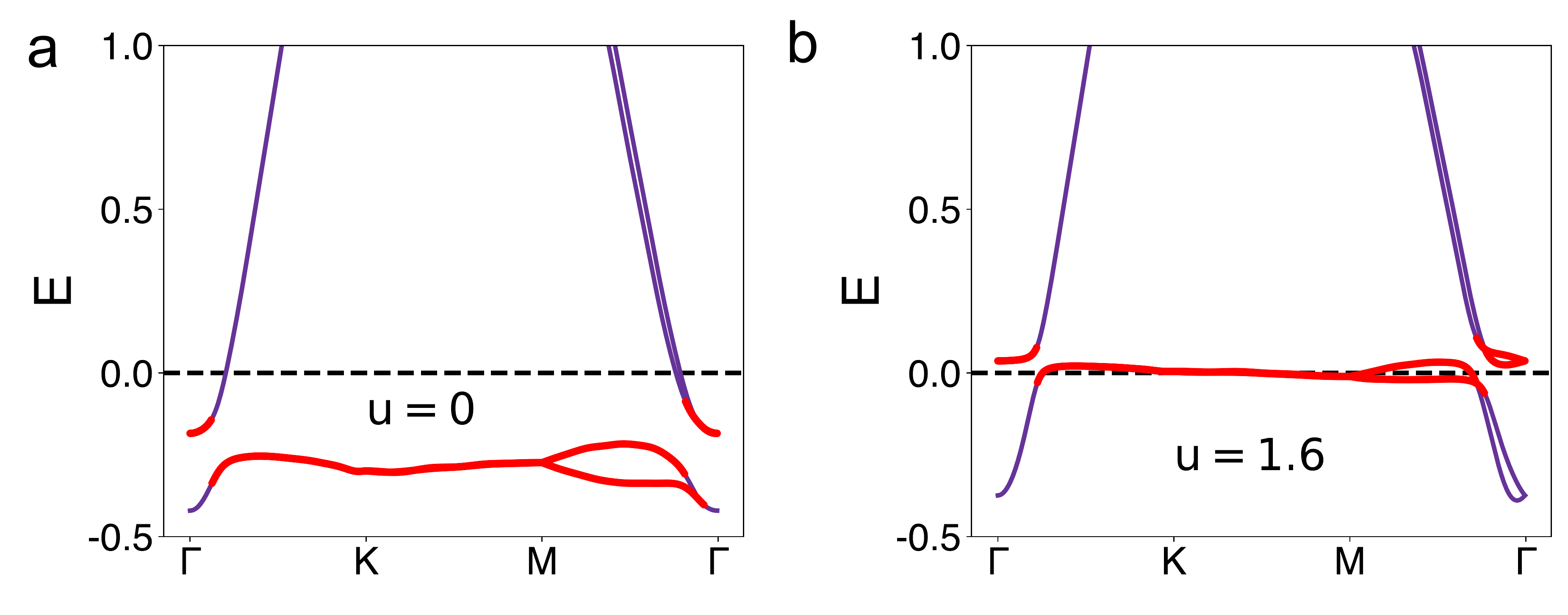}
\caption{{\bf The electronic dispersion in the presence of the SOC.}
Shown here are the electronic dispersions 
for $u=0$ ({\bf a}) and 
for $u=1.6$ ({\bf b}), with the SOC in the original lattice to have a strength of $t_{soc}=0.06$. 
}
\label{fig:bd_soc}
\end{figure}

As shown in Fig.\,\ref{fig:bd_soc}, the SOC generically gaps out the Dirac nodes (located  along $\Gamma-K$) for our 2D model,
both for the cases with and without interactions. 
This leads to flat $Z_2$ topological bands.
We stress that, the flatness of the emergent bands
close to the Fermi energy is robust against the SOC.  

 %%%%%%%%%%%%%

\section*{Supplementary Note 5: Residual interactions and their competition}\label{sec:res_int}
% \subsection*{Residual interactions and their competition}
% \section{Residual interactions and their competition}\label{sec:res_int}

We have so far considered the dominant terms of the interactions that are projected into the Wannier basis: the on-site Coulomb interaction among the most localized molecular (Wannier) orbitals ($d$) and the other on-site interactions between the $d$ orbitals and those of the more extended molecular orbitals ($c_1$ and $c_2$), as described by Eq.\,1 in the main text.
Taking into account these interactions at the saddle-point level of the effective Hamiltonian in the slave-spin approach leads to the emergent flat band, in which the (renormalized) electronic quasiparticle states are primarily associated with the effective $d$ orbitals.

The residual interactions for the emergent flat bands are mainly of three types,
all of which are primarily pertinent to the 
$d$ electrons (as opposed to the $c_1$ and $c_2$ electrons).
The first type is the Ruderman–Kittel–Kasuya–Yosida (RKKY) interaction between the $d$ electrons, mediated by the conduction $c$ electrons. For the generic case with typical fillings of the conduction $c$ electrons, the RKKY interaction tends 
to be antiferromagnetic. However, when the $c$ electrons are dilute, the RKKY interaction would be ferromagnetic.

The second type is the superexchange interaction between the $d$ electrons. This arises because, in the effective (projected) model, the hybridization between the $d$ orbitals in different unit cells is non-zero and so is the hybridization between the $d$ and $c$ orbitals. These kinetic terms induce a coupling between the low-energy renormalized (i.e., coherent) part of the $d$-electron spectral weight near the Fermi energy and the incoherent high-energy part of the $d$-electron spectral weight (which is controlled by both the $u$-interaction and bare $d$-level); when the latter are integrated out, the superexchange interaction ensues. 
The superexchange interaction is typically antiferromagnetic and is maximized near the (orbital-selective) Mott transition point. Deep in a Mott insulating phase, the value of the superexchange interaction can be derived using a direct perturbation theory in terms of the $d$-$d$ hopping and $d$-$c$ hybridization. In the bad metal region, i.e. in the regime of our interest--where the coherent part of the $d$-electron spectral weight is nonzero but small, such that the system is in proximity to the orbital-selective Mott transition -- the superexchange interaction can be constructed by considering the fluctuations beyond the saddle-point level such that the incoherent part of the slave-spin spectral weight is taken into account: This method is in parallel to the derivation for the superexchange interaction in a closely-related slave-particle representation, viz. the slave-rotor approach (see Fig.\,2b of Ref.~\citenum{Ding19.2}).

The third type of residual interactions consists of the direct exchange among the $d$-electrons.
They can be constructed from the projection of the original Hubbard interaction to the Wannier basis.
Although the emergent molecular orbitals are exponentially localized, even the most compact ($d$) orbital still extends over several lattice sites. The shared sites between the nearby molecular orbitals contribute to this ferromagnetic direct exchange interaction.
These typically ferromagnetic interactions between the neighboring sites have the form of the Hund's-like two-body interactions.

We have so far focused on spin exchange interactions. The processes leading to the first and third types of residual interactions naturally lead to density-density interactions between neighboring sites as well.

In summary, the forms of the residual interactions are of the spin-spin and density-density interactions among neighboring sites. An appropriate study of the dominant residual
interactions should 
include the above terms, and the competition between them could lead to different phases.
For example, if the antiferromagnetic interactions prevail over the ferromagnetic terms, they
promote antiferromagnetic-ordering tendencies and the associated 
 quantum criticality. If the ferromagnetic interactions dominate, they may split the Kramers degeneracy, leading to bands with a non-zero Chern number. With the help of other short-range interactions, including the short-range density-density interactions, this regime
 has the potential to yield fractional
 Chern insulating (FCI) states.

\section*{Supplementary Note 6: Slave spin method}
\label{Sec:SS}

In this section, we 
derive the slave spin equations we used in the main text. In the slave spin method, the creation operator for the correlated orbital is re-expressed as the product of a SS operator and an auxiliary fermionic operator: $d_{\sigma}^{\dagger} = f^{\dagger}_{i\sigma}o^{\dagger}_{i\sigma}$, where $o_{i\sigma}^{\dagger} = P^{+}_{i\sigma} S^{+}_{i\sigma} P^-_{i\sigma}$ and $P^{\pm}_{i\sigma} = \frac{1}{\sqrt{1/2 \pm S^{z}_{i\sigma}}}$. To remove the unphysical Hilbert space, the SS and the auxiliary fermion need to follow the constraint:
\begin{equation}
\label{eq:cons}
    S^{z}_{i\sigma} + \frac{1}{2} = n^{f}_{i\sigma}.
\end{equation}
The interaction term can be expressed by the SS operators as described by Eq.\,6 in the main text. The noninteracting part of the Hamiltonian now takes the following form:
\begin{equation}
\begin{aligned}
    H_{0} &= \sum_{ij\sigma} t_{ij} \left(f_{i\sigma}^{\dagger}o_{i\sigma}^{\dagger}o_{j\sigma} f_{j\sigma} + h.c.  \right) - \sum_{i} \mu f_{i\sigma}^{\dagger}f_{i\sigma} \\
    & + \sum_{ij,\alpha\sigma} V^{\alpha}_{ij} \left(f_{i\sigma}^{\dagger}o_{i\sigma}^{\dagger}c_{j\alpha\sigma} + h.c. \right)\\
    & + \sum_{ij,\alpha\beta,\sigma}t_{ij}^{\alpha\beta} \left( c_{i\alpha\sigma}^{\dagger}c_{j\beta} + h.c. \right) + \sum_{i\alpha\sigma} (\Delta_{\alpha} - \mu) c_{i\alpha\sigma}^{\dagger}c_{i\alpha\sigma} \, .
\end{aligned}
\end{equation}
We decouple the $f$ and $o$ 
at the saddle-point level with $ f_{i\sigma}^{\dagger}o_{i\sigma}^{\dagger}o_{j\sigma} f_{j\sigma}    = f_{i\sigma}^{\dagger}f_{j\sigma} \langle o_{i\sigma}^{\dagger} o_{j\sigma} \rangle + \langle f_{i\sigma}^{\dagger}f_{j\sigma}  \rangle o_{i\sigma}^{\dagger}o_{j\sigma} +const. $ and $f_{i\sigma}^{\dagger}o_{i\sigma}^{\dagger}c_{j\alpha\sigma} =  \langle f_{i\sigma}^{\dagger}c_{j\alpha\sigma} \rangle o^{\dagger}_{i\sigma} + \langle o^{\dagger}_{i\sigma} \rangle f_{i\sigma}^{\dagger}c_{j\alpha\sigma} + const$. We 
take the single site decoupling 
 $\langle o_{i} o_{j}^{\dagger} \rangle \approx \langle  o_{i} \rangle\langle  o_{j}^{\dagger} \rangle$, and Taylor expand 
$o^{\dagger}_{i\sigma}$ at the saddle point as follows:
\begin{equation}
\begin{aligned}
    o^{\dagger}_{\sigma} & \approx \langle P^{+}_{\sigma}\rangle S^{+} \langle P^{-}_{\sigma} \rangle + \langle P^{+}_{\sigma} \rangle \langle S^+_{\sigma} \rangle \langle P^{-}_{\sigma} \rangle \frac{1}{2}\left( S^{z}_{\sigma} - \langle S^z_{\sigma}\rangle \right) \left( \frac{-1}{n_{\sigma}}+\frac{1}{1 - n_{\sigma}} \right) \\
    & = O^{\dagger}_{\sigma} + \langle O^{\dagger}_{\sigma} \rangle \frac{1}{2} \frac{n_{\sigma}-1/2}{(1-n_{\sigma}) n_{\sigma} } \left( 2S^z_{\sigma} - (2n_{\sigma} - 1) \right) \\
    & = O^{\dagger}_{\sigma} + \langle O^{\dagger}_{\sigma} \rangle \eta_{\sigma} \left[ 2S^z_{\sigma} - (2n_{\sigma} - 1) \right] \, ,
\end{aligned}
\end{equation}
where 
\begin{equation}
\begin{aligned}
    O_{\sigma}^{\dagger} & = \langle P^{+}_{\sigma} \rangle S^{+}_{\sigma} \langle P^{-}_{\sigma} \rangle \\
    \eta_{\sigma} & = \frac{1}{2} \frac{n_{\sigma} -1/2}{(1-n_{\sigma})n_{\sigma} } \, .
\end{aligned}
\end{equation}
We notice that $\langle o_{\sigma} \rangle  = \langle O_{\sigma} \rangle$. Combining the above equations and introducing the Lagrangian multipler $\lambda_{\sigma}$ to enforce the constraint in Eq.~\ref{eq:cons}, we obtain the self-consistent equations as described in Eq.\,5 of the main text. 

The Green's function of the $d$ electron is obtained by the convolution of the SS and the 
auxiliary fermion propagators, with 
\begin{equation}
    G(\kbf, i\omega_{n}) = \sum_{m} G_{S} (i\Omega_m) G_{f}(\kbf,i\omega_n - i\Omega_m) \, .
\end{equation}
After an analytical continuation, we have the spectral function
\begin{equation}
    \begin{aligned}
        A_d(\kbf, \omega) &= \frac{2\pi}{Z} \sum_{\sigma} \sum_{n,m} \langle n| o_{\sigma}|m\rangle \langle m|o_{\sigma}^{\dagger} |n \rangle \delta(\omega - E_{nm} - \epsilon_{\lambda\kbf}) |\Lambda_{\lambda\kbf}^{f}|^2 \\
    & \times \left[ e^{-\beta E_{m}} (1-f_{\lambda\kbf}) + e^{-\beta E_{n}}f_{\lambda \kbf}\right] \, ,
    \end{aligned}
\end{equation}
where $\lambda$ labels the bands solved from $H^{f}$, and $\Lambda_{\lambda\kbf}^{f}$ represents the corresponding $f$-component in the eigenvectors. In addition, $n,m$ denote the eigenvectors of $H^S$, and $E_{nm}=E_{n}-E_{m}$ is the energy difference between the two slave spin states. Finally, $f_{\lambda\kbf}=\frac{1}{e^{\beta\epsilon_{\lambda\kbf }}+1}$ is the Fermi-Dirac distribution function.

\section*{Supplementary Note 7: Evolution of the spectral functions across the orbital-selective Mott transition}
\label{Sec:dos_mott}

In this section, we provide further details on the  evolution of the single-electron spectral weight when the interactions further increase from the case of $u=1.6$ shown in Fig.\,2b of the main text, approaching and into the OSMP. The evolution is primarily seen in the $d$-electron DOS (red lines). The $c$-electron DOS (marked in purple) is not markedly changed.

\begin{figure}[h]
\centering
\includegraphics[width=0.9\columnwidth]{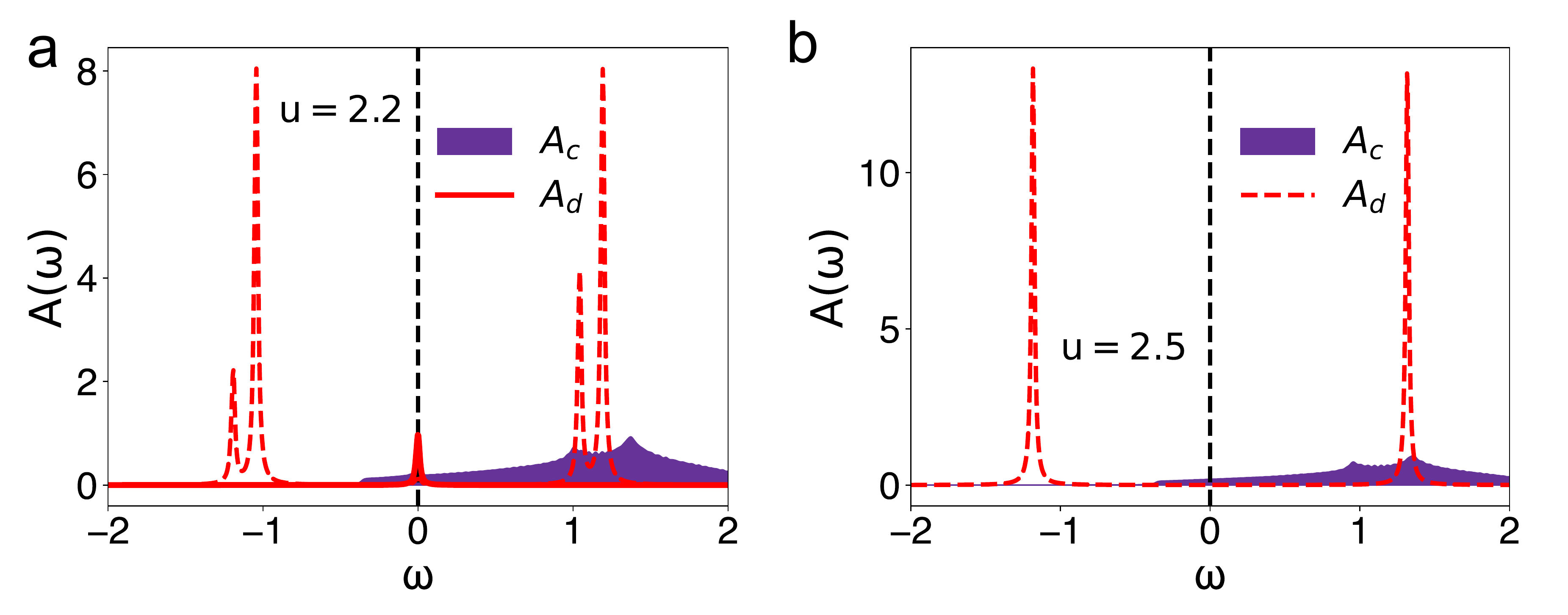}
\caption{
{\bf Evolution of the DOS across the orbital-selective Mott transition.}
{\bf a,}
The DOS of the $d$ and $c$ electrons for 
$u=2.2$, which is just below the critical value 
$u_c=2.4$. The coherent part of the $d$-electron spectrum corresponds to the central peak (the red solid line), and its incoherent part appears as 
side peaks (the red dashed lines). The $c$-electron spectral function is shown in purple.
{\bf b,}
The case of $u=2.5$, which is larger
than  $u_c$ and, thus, the system lies 
inside the OSMP. The $d$-electrons are in a Mott insulating state, and their spectrum contains only the lower and upper Hubbard bands. 
}
\label{fig:dos_mott}
\end{figure}

Fig.\,\ref{fig:dos_mott}a shows the DOS for $u=2.2$, which is smaller than but close to 
the critical value ($u_c=2.4$) for the orbital-selective Mott transition. Compared to the case of 
$u=1.6$ shown in Fig.\,2b of the main text, 
 there is a significant transfer of the 
 $d$-electron spectral weight from the coherent part 
 (the central red peak near the Fermi energy) to the incoherent part (the side peaks marked by the red dashed lines).

Fig.\,\ref{fig:dos_mott}b displays the DOS for $u=2.5$, where the system is in the OSMP (at $u>u_c=2.4$). Here, for the $d$-electron DOS, the coherent peak has vanished. What is left is the incoherent part that comprises 
the  lower and upper Hubbard bands marked by the red dashed lines. 

The slave spin method 
incorporates the nonlocal spin correlations
in the system~\cite{Hu.22.2}, which is important in capturing the orbital-selective Mott transition.
The result here has some similarities with its counterpart arising in the context of the Fe-based superconductors \cite{Huang2022,Yu2021}.
There is however one important distinction: the existence of the 
flat band in the noninteracting dispersion means that the difference between the bandwidths ($D_{\rm flat}$ and $D_{\rm wide}$) is especially large here.

 %%%%%%%%%%%%%
%%%%%%%%%%%%%%%%%%

\section*{Supplementary Note 8: Phase diagram of the effective Multi-orbital Model}
\label{Sec:Phsedig}
\begin{figure}[h]
\centering
\includegraphics[width=1.0\columnwidth]{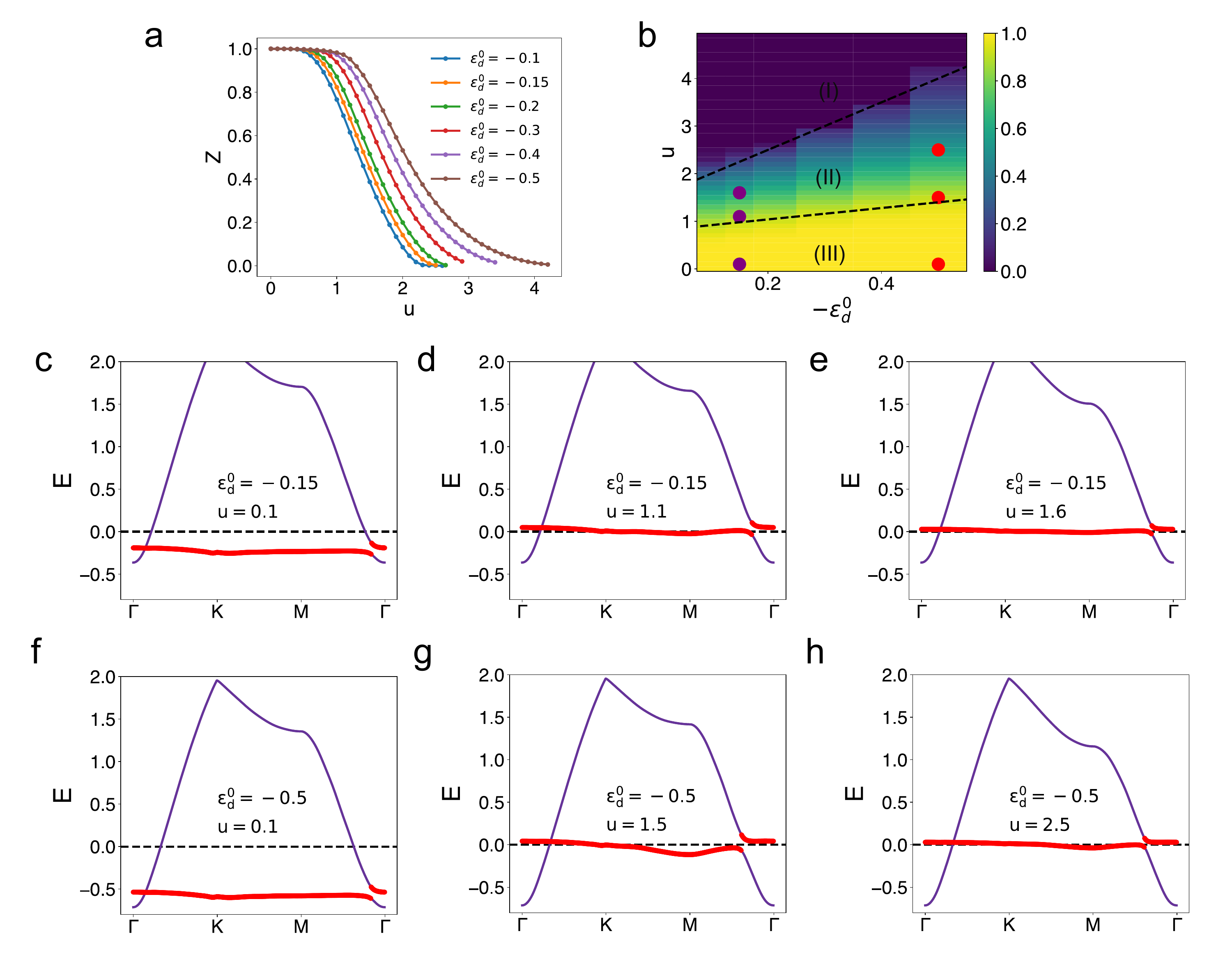}
\caption{
\textbf{Phase diagram for the regime of emergent flat bands.}
{\bf a,} The quasiparticle weight of the more strongly correlated species of electrons ($d$) versus the interaction strength and for various $\epsilon_d^0$. {\bf b,}
The corresponding colormap plot. 
In the parameter setting between the two black lines (region ``II"), an emergent flat band develops in the immediate vicinity of the Fermi energy.
{\bf c-e,}
The dispersion of the coherent single-electron excitations are shown 
%at $u=0.1, 1,4, 2.5$, respectively. 
for parameters corresponding to the purple points marked in {\bf b}. 
{\bf f-h.}
The counterparts for the 
red points marked in {\bf b}.}
\label{fig:zall}
\end{figure}

In this section, we discuss the overall phase diagram in terms of $u$ and the bare local energy level of the $d$ electron ($\epsilon_d^0$).
%on the 
%%global
%overall phase diagram. 
According to the kinetic Hamiltonian as described by Eq.\,4 in the Methods, $\epsilon_d^0=-\mu$. As $\epsilon_d^0$ decreases, the flat band 
further deviates from the Fermi energy. We 
scan the phase diagram as a function of both $\epsilon_d^0$ and the interaction strength $u$. The curves of quasiparticle weight versus $u$ for the $d$ electron are shown in Fig.\,\ref{fig:zall}a for different values of $\epsilon_d^0$, accompanied by the corresponding colormap shown in Fig.\,\ref{fig:zall}b. The 
%global 
phase diagram can be divided into three regimes separated by the black dashed line, as depicted in Fig.\,\ref{fig:zall}b. In the dark blue region (region ``I"), the quasiparticle weight of the $d$ electron completely vanishes, leading to the full localization of the 
$d$ orbital. In the bright yellow region (region ``III"), the quasiparticle weight of the $d$ electron almost equals $1$, and the flat band remains distant from the Fermi energy. The intermediate region (region ``II"), marked as gradually changing shades of green, corresponds to the orbital-selective correlation region, where 
we realize an emergent flat band pinned to the Fermi energy.
This can be further elucidated by comparing the single-particle excitations with a fixed $\epsilon_d^0$ and varying the interaction strengths $u$. The dispersions displayed in Fig.\,\ref{fig:zall}c-e correspond to the parameter settings marked with purple dots, as shown in Fig.\,\ref{fig:zall}b, with $\epsilon_d^0=-0.15$ and $u=0.1$, $1.1$ and $1.5$, respectively. These %represent 
correspond to the weakly interacting region, mixed-valence region, and the Kondo limit. The 
%comparing of the 
dispersions with a local energy further away from the Fermi energy ($\epsilon_d^0=-0.5$) is shown in 
Fig.\,\ref{fig:zall}f-h, displaying a similar trend.

 %%%%%%%%%%%%%
%%%%%%%%%%%%%%%%%%

\section*{Supplementary Note 9: Orbital selective correlations 
in a general setting}
\label{Sec:kagome}
\begin{figure}[h]
\centering
\includegraphics[width=1.0\columnwidth]{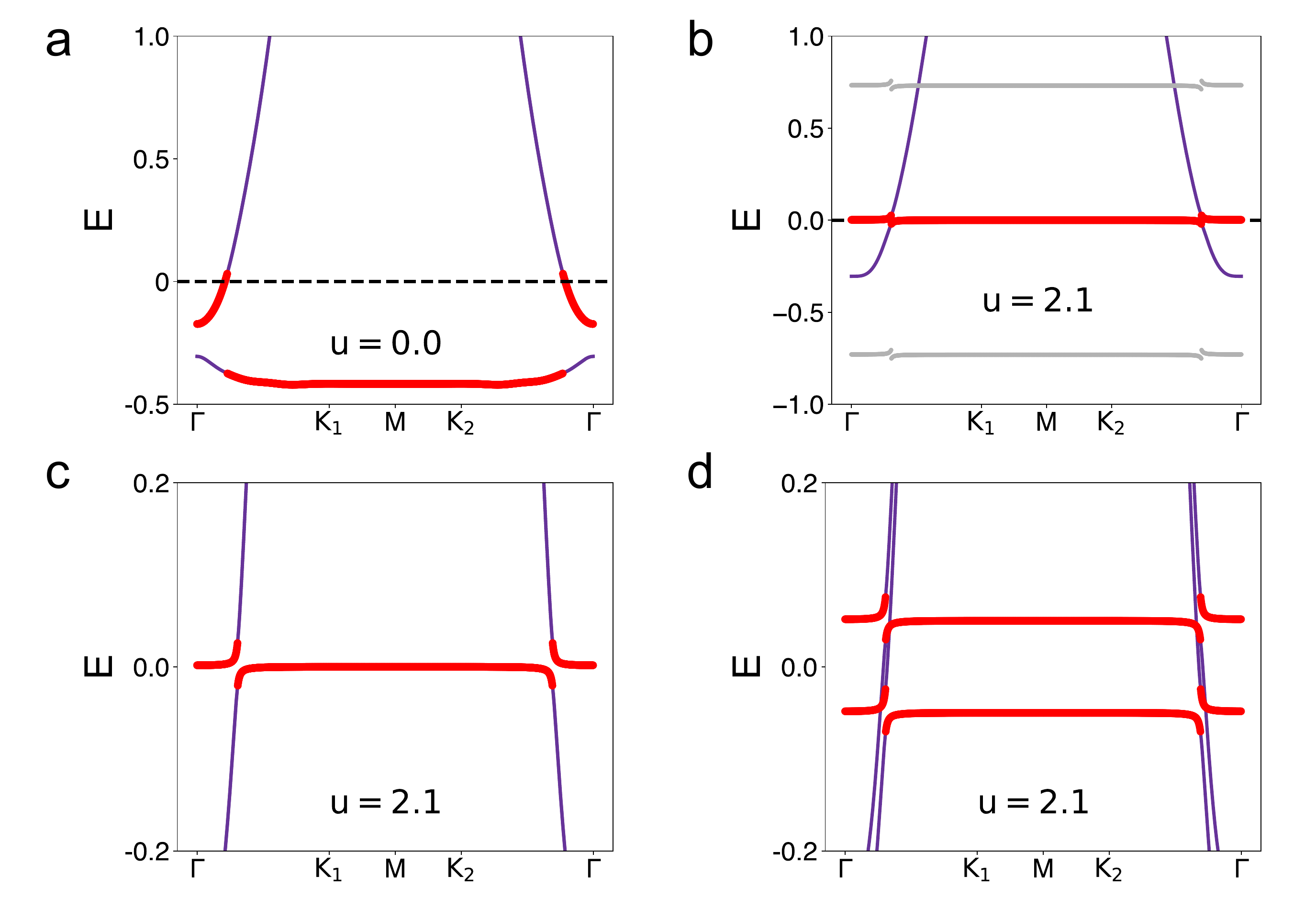}
\caption{ \textbf{Topological Kondo effect in a general setting.}
{\bf a,} The noninteracting band structure. 
{\bf b,} The dispersion of the coherent single-electron 
excitations at 
$u=2.1$. The red solid curve denotes the emergent flat band close to the Fermi energy. The grey lines mark the incoherent single-electron excitations. 
{\bf c,} The zoomed-in view of the emergent flat band. 
{\bf d,} The band structure at $u=2.1$ with a Zeeman splitting $m_z=0.05$.
}
\label{fig:kagome}
\end{figure}

In this section, we illustrate that the conclusion we have reached is applicable to other types of geometry-induced flat band systems. 
(For an
overall discussion, see the main text.) The example presented here is a model defined on the kagome lattice, which has recently been found to display a Kondo-destruction quantum critical point~\cite{Chen2023-kagome}. Here,
we consider the case when the energy dispersion in the non-interacting limit is shown in Fig.\,\ref{fig:kagome}a, where the flat band is situated away from the Fermi energy. The dispersion for the interacting case ($u=2.1$) is displayed in Fig.\,\ref{fig:kagome}b, where once again, we observe the emergence of a flat band close to the Fermi energy. Due to the inclusion of spin-orbital coupling, the obtained solution is a topological insulator (TI), featuring a hybridization gap as depicted in the zoomed-in plot in Fig.\,\ref{fig:kagome}c. 
Like in the canonical Kondo systems,
the system is highly tunable.
As shown in Fig.\,\ref{fig:kagome}d, a relatively small Zeeman field leads to a substantial change in topology, where the TI gap is closed and nodal lines appear, which are protected by the $M_z$ symmetry.

 %%%%%%%%%%%%%
%%%%%%%%%%%%%%%%%%

\end{document}